\documentclass[showpacs,aps,floatfix] {revtex4}
\usepackage{pst-all}
\newcounter{subequation}[equation]
\makeatletter \fussy
          
   \def \L  {\Lambda } \def \l {\lambda }      \def \T {\Theta }  \def \a {\alpha } \def \dh {\partial } \def \d
{\delta } \def \D {\Delta }   \def \g
{\gamma }   \def \o {\omega }  \def \S {\Sigma } \def \s {\sigma }                \def  \ud { {1 \over 2} }    \def \td {{3 \over  2 }} 
\def \cd {{5 \over  2 }}         \def
\cddd { {\cal D } }   \def
\calp { {\cal P } }      \def \caln { {\cal N } } 
     \def \cald { {\cal D } }       
      
 \def \Eslash {E \kern-.5em\slash} \def \pslash
{p \kern-.5em\slash} \def \kslash {k \kern-.5em\slash} \def \Dslash {D
\kern-.5em\slash} \def \hslash {h \kern-.5em\slash} \def \dslash {\partial
\kern-.5em\slash} \def \vslash {v \kern-.5em\slash}

    \def\NPB#1#2#3 {{\rm Nucl.~Phys.}  {\bf{B#1}},
{#3} (#2)} \def\NPA#1#2#3 {{\rm Nucl.~Phys.}  {\bf{A #1}}, {#3} (#2)}
\def\PLB#1#2#3 {{\rm Phys.~Lett.}  {\bf{B #1}}, {#3} (#2)} \def\PR#1#2#3
{{\rm Phys.~Rep.}  {\bf#1}, {#3} (#2)}  \def\PRD#1#2#3 {{\rm Phys.~Rev.}
{\bf{D #1}}, {#3} (#2)}  \def\PRL#1#2#3 {{\rm Phys.~Rev.~Lett.}  {\bf{#1}},
{#3} (#2)}  \def\ZPC#1#2#3 {{\rm Z.~Phys.}  {\bf C #1}, {#3} (#2)}
\def\JHEP#1#2#3 {{\rm JHEP} {\bf C#1}, {#3}  (#2)}      \def\IJMP#1#2#3
{{\rm Int. J. Mod. Phys.}  {\bf A #1}, {#3} (#2)}  \def\JMP#1#2#3 {{\rm
J. Math. Phys.}  {\bf #1}, {#3} (#2)} \def\PTP#1#2#3 {{\rm
Prog. Theor. Phys.}  {\bf{#1}}, {#3} (#2)} \def\FP#1#2#3  {{\rm
Fortsch. Phys.}  {\bf{#1}}, {#3} (#2)} \def\CMP#1#2#3  {{\rm
Comm. Math. Phys.}  {\bf{#1}}, {#3} (#2)}

\def\thesubequation{\theequation\@alph\c@subequation}
\def\@subeqnnum{{\rm (\thesubequation)}}
\def\slabel#1{\@bsphack\if@filesw {\let\thepage\relax
\xdef\@gtempa{\write\@auxout{\string
\newlabel{#1}{{\thesubequation}{\thepage}}}}}\@gtempa \if@nobreak
\ifvmode\nobreak\fi\fi\fi\@esphack}
\def\subeqnarray{\stepcounter{equation}
\let\@currentlabel=\theequation\global\c@subequation\@ne
\global\@eqnswtrue
\global\@eqcnt\z@\tabskip\@centering\let\\=\@subeqncr
$$\halign to \displaywidth\bgroup\@eqnsel\hskip\@centering
$\displaystyle\tabskip\z@{##}$&\global\@eqcnt\@ne
\hskip 2\arraycolsep \hfil${##}$\hfil &\global\@eqcnt\tw@ \hskip
2\arraycolsep $\displaystyle\tabskip\z@{##}$\hfil
\tabskip\@centering&\llap{##}\tabskip\z@\cr}
\def\endsubeqnarray{\@@subeqncr\egroup
	     $$\global\@ignoretrue}
\def\@subeqncr{{\ifnum0=`}\fi\@ifstar{\global\@eqpen\@M
\@ysubeqncr}{\global\@eqpen\interdisplaylinepenalty \@ysubeqncr}}
\def\@ysubeqncr{\@ifnextchar [{\@xsubeqncr}{\@xsubeqncr[\z@]}}
\def\@xsubeqncr[#1]{\ifnum0=`{\fi}\@@subeqncr
\noalign{\penalty\@eqpen\vskip\jot\vskip #1\relax}}
\def\@@subeqncr{\let\@tempa\relax \ifcase\@eqcnt \def\@tempa{& & &}\or
\def\@tempa{& &} \else \def\@tempa{&}\fi \@tempa
\if@eqnsw\@subeqnnum\refstepcounter{subequation}\fi
\global\@eqnswtrue\global\@eqcnt\z@\cr} \let\@ssubeqncr=\@subeqncr
\@namedef{subeqnarray*}{\def\@subeqncr{\nonumber\@ssubeqncr}\subeqnarray}
\@namedef{endsubeqnarray*}{\global\advance\c@equation\m@ne%
\nonumber\endsubeqnarray}

\@addtoreset{equation}{section}
\renewcommand{\theequation}{\thesection.\arabic{equation}}

 \newcommand{\be}{\begin{equation}}
\newcommand{\ee}{\end{equation}} \newcommand{\ba}{\begin{array}}
\newcommand{\ea}{\end{array}} \newcommand{\bea}{\begin{eqnarray}}
\newcommand{\eea}{\end{eqnarray}}
\newcommand{\bsea}{\begin{subeqnarray}}
\newcommand{\esea}{\end{subeqnarray}} 
  
\begin{document}
%\begin{titlepage}
%\draft
%\wideabs{
\rightline{IPhT: t09/154}
\rightline{TUM-HEP: 09/737}
\rightline{arXiv:0909.4497}
% Classe REVTEX4 
\title{Searching singlet extensions of the   supersymmetric standard
  model  in  $ Z_{6-II}$ orbifold
compactification  of heterotic string} \thanks {\it Supported by
the Laboratoire de la Direction des Sciences de la Mati\`ere du
Commissariat \`a l'Energie Atomique }  \author{M. Chemtob} \email{marc.chemtob@cea.fr}
\affiliation{CEA, DSM, Institut de Physique Th\'eorique, IPhT, CNRS, MPPU,
URA2306, Saclay, F-91191 Gif-sur-Yvette, France}  \author{P. Hosteins}
\email{pierre.hosteins@ph.tum.de} 
\affiliation{Physik-Department T30, Technische Universit\"at
  M\"unchen, \\   
James-Franck-Stra\ss e, 85748 Garching, Germany} 
\date{\today} 
\pacs{12.60.-i,12.60.Jv}

% 11.25.Mj 	Compactification and four-dimensional models
% 11.25.Wx 	String and brane phenomenology
% 12.60.-i 	Models beyond the standard model
% 12.60.Jv 	Supersymmetric models
%%%%%%%%%%%%%%%%%%%%%%%%%%%%%%%%%%%%%%%%%%%%%  
%\tableofcontents
%\addtocontents{toc}{\protect\setcounter{tocdepth}{2}}

\begin{abstract}  
We search for supersymmetric standard model realizations with extra
singlets and extra $ U(1)$ using the heterotic string compactification
on the $ Z_{6-II}$ orbifold with two Wilson lines.  The effective
superpotential produced through the vacuum restabilization mechanism
is examined for three representative Pati-Salam string models obtained
in the literature.  An automated selection of semi-realistic vacua
along flat directions in the non-Abelian singlet modes field space is
performed by requiring the presence of massless pairs of electroweak
Higgs bosons having trilinear superpotential couplings with massless
singlet modes and the decoupling of color triplet exotic modes needed
to suppress $B $ and $ L $ number violating processes.
\end{abstract} \maketitle

%\renewcommand{\thefootnote}{\alph{footnote}}
%\vspace{-.8cm}
%\tableofcontents
%\addtocontents{toc}{\protect\setcounter{tocdepth}{2}} 
%\vfill\eject

\section{Introduction}
\label{sec:intro}
Model building with the heterotic
string~\cite{Font:1989aj,Aldazabal:1995yw,Aldazabal:1995cf,Katsuki:1989bf,erlerklemm,dixon1,giedt02,Forste:2005gc,Forste:2005rs}
has received a vigorous stimulus thanks to the recent
focus~\cite{Kobayashi:2004ya,Kobayashi:2006wq,Kobayashi:2004ud,Buchmuller:2005sh,Buchmuller:2006ik,Buchmuller:2005jr,Lebedev:2006tr,Lebedev:2006kn,Forste:2006wq,forste06,Lebedev:2007hv,Ratz:2007my,raby0705,raby0710,raby0805,buchdt07,buch08}
on anisotropic compactifications on the orbifold $ Z_{6-II}$. Recent
reviews are given in~\cite{ramos09,vaude09}. Satisfactory representative 
vacuum solutions   are also   reported   for the   $ Z_{12-I} $
orbifold~\cite{Kim:2006hw,Kim:2006zw,Kim:2007mt,Kim:2007fa,kim07}.
The exploration  of the vast moduli space of the $ Z_{6-II}$ orbifold with  two  authorized
Wilson lines~\cite{Lebedev:2006kn,Lebedev:2007hv}  was found to give  access to a fertile mini-landscape
of vacua for minimal supersymmetric standard models descending from
5-d or 6-d grand unified theories.  Sampling   the
still wider  space of  solutions   with three Wilson
lines~\cite{lebedev08}  leads to  equally   hopeful  conclusions. 

 One characteristic feature of these top-down constructions is the profusion of $U(1)$ gauge factors
which remain after the first gauge symmetry breaking from the orbifold
gauge twist and the presence of Wilson lines. One of them is
anomalous, creating a one-loop Fayet-Iliopoulos D-term a bit below the
string scale that needs to be cancelled. This step usually leaves a
lot of freedom due to the large vacuum degeneracy. In the vacuum
restabilisation process, the $U(1)$ gauge symmetries will get broken
and most  of the extra modes beyond the MSSM (minimal supersymmetric
standard model) matter content, but hopefully not all of them, will
decouple with large masses.  Except for searches of extensions
including right-handed neutrino
supermultiplets~\cite{giedt05,lebedev07}, most studies have restricted
to solutions in which the whole set of Standard Model singlet modes
acquire large masses and decouple from the low energy field theory. 
It is natural, however, to ask whether the large moduli space of these
compactifications does not include regions where extra singlets and
extra $ U(1)$ symmetries might realize the NMSSM (next-to-minimal
supersymmetric standard model) or one of the related versions proposed
in the literature~\cite{cvetic96,cvetic97,erler02,han04,demir05,lee07}
and analyzed over the years in
phenomenological~\cite{ellis89,Barger:2006dh,Barger:2006sk,chiang08,langvin08,langacker08}
and formal~\cite{Bagger95,abel95,pana99,pilaftsis02,pana02} studies.
(After completion of the present work, there appeared
a   study of NMSSM  realizations  complying with the  severe 
selection criteria of~\cite{Lebedev:2006kn,Lebedev:2007hv}   
which  reports  that solutions exist  only  in the case  with three Wilson
lines~\cite{sanchez10}.) 

 Filling this gap is the purpose of the present work.  
Rather  than pursuing  a statistical exploration  of the moduli space of
heterotic string compactifications on the $ Z_{6-II}$ orbifold, we
restrict consideration in this preliminary study    to 
the three representative 5-d Pati-Salam models,  designated
in~\cite{Kobayashi:2004ya} as models $A1, \ A2, \ B $, and 
apply   techniques developed in past works.   The early
studies along these lines focused on flat directions for the
non-Abelian singlet modes realizing the minimal supersymmetric
standard model~\cite{cleav198,cleav298,cleav398,cleaver99} or the
Pati-Salam model~\cite{leontaris99}, while subsequent studies have
explored field directions along scalar modes charged under the
non-Abelian gauge group
factors~\cite{cleaver02,Faraggi:2006bs,cleaver07,cleaver08}.

 Having in hand the string massless spectra for a set of
representative solutions, we use a Fortran computer program to
implement an automated search of the flat directions which exhibit a
satisfactory supersymmetric effective action.  Our study is somewhat
close in spirit to that of~\cite{leontaris99}.  Concentrating on string
theory realizations~\cite{antoleon89,allan96,leon96,dent06} of the
Pati-Salam model~\cite{patisalam73} comes with a number of  well-known
advantages.  We shall  not detail in the present work the   Pati-Salam to Standard Model ($PS\to
SM$) gauge symmetry breaking,  but note that the use of Pati-Salam group
multiplets   merely serves a convenient book-keeping
purpose, since the representation content with respect to the Standard
Model gauge group is then uniquely determined.   

The contents of the present work are organized into four sections.
Section~\ref{sec1} discusses general features of the low energy theory
with Pati-Salam gauge symmetry.  A brief review of the supersymmetry
preserving flat directions in the presence of an anomalous $U(1)_A$
gauge symmetry is provided in Appendix~\ref{appexa}.  In order to
clarify certain fine points that have caused us much confusion, we
found useful to include in Appendix~\ref{appexb} a brief review of the
string theory construction for the $ Z_{6-II} $ orbifold complementing
the detailed discussions
in~\cite{Kobayashi:2004ya,Buchmuller:2006ik,Lebedev:2007hv}.  The core
part of the present work figures in Section~\ref{sec2}, where we
present the results of our searches of singlet extended supersymmetric
standard models for the Pati-Salam string models $ A1, \ A2, \ B$
of~\cite{Kobayashi:2004ya}.  A general discussion of results and a
summary of our conclusions figures in Section~\ref{sec3}.

\section{Low energy theory}
\label{sec1}

\subsection{General features of Pati-Salam string models}
Our interest is on the 4-d orbifold compactifications of the
$E_8\times E_8^{'}$ heterotic string satisfying $\caln =1$
supersymmetry, with gauge symmetry group, $ G\times G' $, including a
Pati-Salam observable sector,{ { 
\begin{equation}
G = G_{422} \times \prod _{a=1} ^{N_a} U(1)_a, \qquad [G_{422}= SU(4) \times
SU(2) _L \times SU(2) _R ]
\end{equation}
and a hidden sector,
\begin{equation}
G'= SO(10)'\times SU(2)' \quad\text{ or }\quad G'= SO(10)',
\end{equation}}}
with  the corresponding Abelian factors, 
$ U(1) ^ {N_a} , \ [N_a=5 , 6 ]$.  The gauge group
representations consist of three chiral generations of bifundamental
matter supermultiplets, $ f _i\sim (4,2,1), \ f ^c _i \sim (\bar
4,1,2) $, a number of electroweak Higgs bidoublet supermultiplets, $
h_i \sim (1,2,2)$, and a number of vector bifundamental
supermultiplets, $ f '_i, \ \bar f'_i $ and $ f ^{'c}_i ,\ \bar
f^{'c}_i $, which accomplish the $PS\to SM$  Higgs mechanism for 
the gauge symmetry breaking, $ SU(4) \times
SU(2)_R \to SU(3) _c \times U(1)_Y $, along with a set of non-Abelian
singlet modes, $ \phi _i \sim (1,1,1) (1',1')$.  There also occur
modes with exotic quantum numbers: \be C_i\sim (6,1,1), \ \ d ^r _i
\sim(1,1,2) , \ \ d^l _i \sim (1,2,1) , \ \ q _i\sim (4,1,1), \ \ \bar
q_i\sim (\bar 4,1,1); \ee hidden sector modes: \be \S ' _i \sim (16
',1') ,\ \ \bar \S ' _i \sim (\overline{16'} ,1') ,\ \ T ' _i \sim (10
',1') ,\ \ d ' _i \sim(1',2'); \ee along with modes charged under the
observable and hidden groups: \be h ^{'l} _i \sim (1,2,1)(1',2') ,\ \
h ^{'r} _i \sim (1,1,2) (1',2').  \ee Among the set of non-Abelian
singlets, $ \phi _i $, we distinguish the modes without and with
oscillator excitations, $S_i $ and $ Y_i$.  We group the singlets into
two subsets, the first including the fields $ \hat \phi _i $ excited
along the flat directions with large vacuum expectation values (VEVs)
of the order of the string or Planck mass scales, $m_s$ or $M_\star $,
and the second including the dynamical fields, $\tilde \phi _i $ and $
\phi _i $, with undetermined or vanishing VEVs and masses.
 
The $ PS \to SM $ phase transition is assumed to occur at a mass scale
$ m_{PS} $ near the string scale.  This hypothesis is in harmony with
the renormalization group analysis of the gauge coupling
constants~\cite{Kobayashi:2004ya} and with related phenomenological
studies~\cite{allan96,leon96,dent06}.  At the $ PS \to SM $ gauge
symmetry breaking, the above listed multiplets decompose into Standard Model group $ G_{321} =
SU(3) \times SU(2) _L \times U(1)_Y $ multiplets as, \bea && f \sim
(4,2,1) \to q+l = (u \ d ) + (\nu \ e ) \sim (3,2)_{1\over 6 } +
(1,2)_{-\ud } , \cr && f^c \sim (\bar 4,1,2) \to q^c + l^c = (d^c \
u^c ) + ( e^c \ \nu ^c ) \sim (\bar 3,1)_{2\over 6 } + (\bar 3,1)_{-
{4\over 6 } } + (1,1)_{1} + (1,1)_{0}, \cr && h \sim (1,2,2) \to
H_u+H_d \sim (1,2)_{\ud } + (1,2)_{-\ud } ,
%,\cr && \bar f ^c \sim (4,1,2) \to \bar q^c + \bar l^c = (\bar d^c \
%\bar u^c ) + (\bar e^c \ \bar \nu ^c ) \sim (3,1)_{- {2\over 6} } +
%(3,1)_{4\over 6 } + (1,1)_{0} + (1,1)_{-1}  , \ 
\cr && C \sim (6,1,1) \to \bar g^c + g^c \sim (3,1)_{- {1\over 3} } +
(\bar 3,1)_{1\over 3} , \cr && q \sim (4,1,1) \to g' + E =
(3,1)_{1\over 6} + (1,1) _{-\ud } ,\
%\bar q =(\bar 4,1,1) \to (\bar 3,1)_{- {1\over 6 } } + (1,1) _{\ud }
%,  
\cr && d^l \sim (1,2,1) \to (N^l \ E^l) \sim (1,2) _0 , \cr && d^r
\sim (1,1,2) \to E^r + N^r \sim (1,1) _{-\ud } + (1,1) _{\ud }.  \eea
Analogous decompositions hold for the conjugate representations, $\bar
f \sim (\bar 4,2,1), \ \bar f ^c \sim (4,1,2), \ \bar q \sim (\bar
4,1,1).$
%  The exotic color triplet modes  hat we  denote by $g$  are  denoted
%  by   KRZ  as $D$. 
The superpotential couplings of Pati-Salam modes of lowest order
decompose into couplings between SM modes as 
{  {\bea 
h_kh_l &=& H_{u,k} H_{d,l} +H_{d,k} H_{u,l}, \cr
q _k \bar q_l &=& g'_k\bar g'_l + E _k\bar E_l, \cr
C_k C_l &=& \bar g ^c_k  g_l^c + \ g_k ^c \bar g_l^c, \cr
hff^c &=& H_u q u^c + H_d q d^c + H_d l e^c , \cr 
C ff &=& \bar g^c  qq, \cr
C f^c f^c &=& g^c  u^c d^c + \bar g ^c u^ce^c+\bar g^c d^c\nu ^c ,\cr
C \bar f^c \bar f^c &=& \bar  g^c \bar  u^c\bar  d^c +  g ^c\bar  u^c\bar e^c+ g^c \bar d^c 
\bar  \nu  ^c .  
\eea}}

\subsection{Search  strategy}

We search vacua of the three chiral generation models $A1, \ A2, \ B $
of~\cite{Kobayashi:2004ya} with the following characteristics at low
energies:
\begin{itemize}
\item One or several non-Abelian gauge group singlets, $\phi _k$,
coupled to one or several pairs of electroweak Higgs boson bidoublets
by trilinear effective superpotential terms, $ h_ih_j \phi _k$;
\item Extra $ U(1)'$ symmetries~\cite{langacker08} possibly broken at
mass scales much lower than $ O(M_\star ) $;
\item A secluded sector with additional singlets~\cite{erler02,lee07}
belonging to an approximate flat direction and charged under the extra
$ U(1)'$ only.  The simplest version includes in addition to $s$ three
singlets~\cite{erler02,lee07} $s_1, \ s_2 , \ s_3 $ belonging to a
flat direction approximately lifted by a small coupling constant $\l '
$ in the effective renormalizable superpotential, $ W _{EFF}= \l H_u
H_d s + \l ' s_1 s_2 s_3 $.
\end{itemize}  
The superpotential is decomposed into three main components, $W_s, \
W_2 ,\ W_3 $, associated to the couplings of the non-Abelian singlet
fields, $ \phi _i $, alone or multiplied by two and three fields
charged under the Pati-Salam group.  Each component is built as a
power expansion in the
$ \phi _i $  of the form   
\bea 
W_s(\phi  _k) &=& \sum _{n} W_s^{(n)} (\phi  _k)  = \sum _{n, m}  
s_m ^{(n)} \prod  _{i,j}   S_i ^{p_i^{(m)} } Y _j
 ^{q_j^{(m)} },   
\cr  W_2 (\phi  _k)&=& r _{ij}  (\phi  _k) f_i ^c \bar  f_j ^c + p_{ij}  (\phi  _k) f_i   \bar
f_j  + \mu _{ij} (\phi  _k)  h_i  h_j   + \tau _{ij}  (\phi  _k) C_i  C_j + \s _{ij}  q_i
\bar q_j ,
\cr W_3 (\phi  _k) &=& C_k  [ c _k ^{ij}  (\phi  _l)  f _i^c  f_j^c 
+  c _k ^{'ij} (\phi  _l) \bar   f _i^c  \bar f_j^c  +  l _k ^{ij} (\phi  _l)   f _i  f_j  + 
e _k ^{ij}   (\phi  _l) q_i  q_j    + e _k ^{'ij}  (\phi  _l)  \bar
q_i \bar  q_j  ] \cr  &&   +   \l _k ^{ij}  (\phi  _l)   
h_k  f_i f^c _j  +\l _k ^{ 'ij}   (\phi
_l)   h_k  \bar f_i \bar  f^c _j     , \cr  &&  \label{eqw2w3}  \eea 
where $ p_i^ {(m)} ,\ q_j^{(m)} \in Z _+ ,\ \sum _{i,j}
  ( p_i^ {(m)} +q_j^{(m)}  ) \geq n $ and the  summations over $ f_i ,\  f^c _i$ (and their
complex  conjugates)    include both  the matter
and Higgs  multiplets. The   coefficient    functions 
$\mu _{ij} (\phi  _k) ,\   \s  _{ij}(\phi  _k),
\ \tau _{ij}  (\phi  _k), \ \cdots $ are    given  by  infinite
power expansions in    the singlet     fields of same form  as  
that  displayed   above for   $ W_s ^{(n)}$. 
 Integrating out the massive  modes $C_i$,   by means of  the classical  fields 
equations,  can produce  baryon and/or lepton number  violating 
couplings  represented  by  local   operators  of dimension
$\cddd = 4$:      $ lqd^c, \ u^cd^cd^c , \ lle^c$, or  of dimension $\cddd = 5$:    
$ qqql ,\ u^c d^c d^c \nu ^c  $.  The  resulting   dangerous  local  operators,
\bea
W_{EFF} &=& ( l_k (\tau  ^{-1}) _{kl}) \   l_l ffff  + ( l _k (\tau  ^{-1}) _{kl}  c_l
) \  fff^cf^c + ( c _k  (\tau  ^{-1})_{kl}  c_l ) \  f^cf^cf^cf^c 
\cr &\to&   {1\over  M_\star }    [ l ^2 (\phi ) qqql + l  (\phi )  c 
 (\phi )  qq  d^cu^c + c ^{2} (\phi ) u^cd^c d^c
\nu ^c  ]   + {m_{PS} \over  M_\star }  c ^2 (\phi ) u^c d^c
d^c  ,   
\eea
may  compete with   similar  couplings already  present at   the
compactification scale.   

Our search strategy consists of four stages.
\begin{itemize}
\item Firstly, we construct (up to some fixed order) the superbasis of
holomorphic invariant monomials \be P_\a ( \phi ) = \prod _i \phi _i ^
{ r^ \a _i } = \prod _i S_i ^ { n^ \a _i } Y_i ^ { m^ \a _i } , \quad
[ r^ \a _i, \ n ^ \a _i, \ m ^ \a _i \in Z_+ ] \ee which solve the D
flatness conditions, $ Q_a (P_\a ) =0, \ Q_A (P_\a )\times \text{sign}
\ ( \xi _A ) < 0 $, where $ Q_a, \ [ a=1, \cdots , N_a -1]$ denote the
anomaly free Abelian charges and $ Q_A$ the single anomalous Abelian
charge with Fayet-Iliopoulos parameter $\xi _A$ defined in
Eqs.~(\ref{eqgs1}) and (\ref{equniv}).
\item Secondly, we construct the superpotential $W_s$ in the
 non-Abelian singlet modes (up to some fixed order) satisfying the
 gauge symmetries and the string selection rules in
 Eqs.~(\ref{eqsel}).
\item Thirdly, we scan the D flat monomials $P _\a (\phi _i) $ and
compare each of these in turn with the allowed couplings of singlets
in $W_s(\phi _i) $, in order to determine which of the D flat
monomials obey  the  type $A$ or $B$  F flatness,  such that 
there are  no allowed  couplings  in the superpotential $W_s$ with 
all field factors, or all but a single field factor, included in $P_\a (\phi _i) $.
(Said otherwise, type $A$ or $B$ lifting of a flat direction $P_\a
(\phi _i) $ occurs whenever  some   superpotential monomial  {  {with only 1 or no field having zero expectation value is allowed}}.)  Our terminology slightly
deviates from the original one in~\cite{cleav198,cleav298,cleaver07}
which referred to the subset of monomials in $W_s$ allowed by the
gauge invariance rules alone.

The conservative approach of exploring the space of  supersymmetric  
vacua  consists in restricting to the type $B$
directions, on the grounds that these preserve local supersymmetry,
thanks to the vanishing scalar potential or cosmological constant, $
<V>=0$, and can be made flat to all orders by imposing a finite number
of conditions.  The existing 
applications~\cite{cleav298,cleaver07,cleav398,leontaris99} typically
stop at the trilinear or fourth orders terms of $ W_s (\phi ) $.  An
alternative procedure is pursued
in~\cite{Buchmuller:2005jr,Lebedev:2006kn,Kim:2007mt} which consists
in selecting a (typically sizeable) set of singlet field directions
providing for a satisfactory effective action and solving next the D
and F flatness equations out to high orders for these singlets.  In
the present work, we follow an intermediate approach in which we
restrict to flat directions and superpotential monomials of low orders
only but include both type $A$ and $B$ flat directions.  Since
supersymmetry must get broken anyway, we argue that it makes sense to
consider the approximate flat directions which are lifted by
non-renormalizable F terms of reasonably high orders.  To motivate our
choice we note that the condition $ <W_s> \approx 0$ is automatically
satisfied for models preserving an approximate R
symmetry~\cite{kapp09}.  We also observe that an initially non F flat
direction can be repaired into a flat one by using the cancellation
with contributions from higher order  and/or 
moduli dependent couplings~\cite{pokorski98}.  One
may tie the order of the F term lifting operator to the supersymmetry
breaking scale $F$ by  considering the case of a single flat direction
$\phi $ lifted by the superpotential, $ W_s ( \phi )= \phi ^{p+3} /
M_\star ^p $.  Assuming that $\phi $ picks up from radiative
corrections a tachyonic soft supersymmetry breaking mass term, $ V (
\phi ) = - m_W ^2 \vert \phi \vert ^2 $, then the scalar potential
minimization, \bea && V ( \phi ) = - m_W ^2 \vert \phi \vert ^2 +
{(p+3)^2 \over M_\star ^{2p} } \vert \phi ^{p+2 }\vert ^2 \
\Longrightarrow \ \vert F \vert ^\ud = V ^{1/4}(\phi _{min}) \simeq
(m_W ^{p+2} M_\star ^{p} ) ^{1\over 2(p+1)} , \eea yields the
following acceptable range for the breaking scale, $ \vert F \vert
^\ud \leq (10^6 \ - \ 10^8) $ GeV, for $ p =(1 \ - \ 3 ) $.  Larger
$p$ yield higher scales.  It should be noted that the F term lifting
by non-renormalizable operators also plays a useful r\^ole in string
theory models realizing gauge symmetry
breaking~\cite{kalara87,faraggi92} or supersymmetry
breaking~\cite{faraggi96} at intermediate scales.

\item Fourthly, we select some supersymmetry preserving flat
directions, identified by sets of excited singlet fields $\hat \phi _i
$, and evaluate the effective superpotential between the non-singlet
modes, while treating the fields $\hat \phi _i $ VEVs as free
parameters.  Our main focus will be on the bilinear and trilinear
couplings of the electroweak Higgs modes, $ h_ih_j (\mu _{ij} (\hat
\phi ) + \l _{ij} ^k (\hat \phi )\phi _k ) $. However, for
completeness, we shall also examine the bilinear couplings of the
modes carrying the bifundamental, sextet and quartet representations,
in order to test the decoupling of mirror generations and exotic modes
and the absence of baryon number violation.
\end{itemize}  
   
We pause briefly at this point to clarify some technical issues.  The
non-Abelian singlet modes VEVs are $ 0 $-dimensional if they are
completely fixed by the gauge constraints, in which case they are
roughly set at, $ \hat \phi = O( ( -\xi _A / Q_A ( \hat \phi ) )^\ud )
$. Otherwise, they may be 1-d or higher if they depend on a single or
several free complex parameters.

A given holomorphic monomial of the non-Abelian singlets, $ P _\a =
\prod _{i=1}^n \phi _i^ {r_i ^\a } $, is D-flat if it carries
vanishing anomaly free charges, $ Q_a (P _\a ) =\sum _{i=1}^n Q_a
(\phi _i ) r_i ^\a =0 $, and a finite anomalous charge, $Q_A ( P _\a )
=\sum _{i=1}^n Q_A (\phi _i )r_i^\a \ne 0 $ of opposite sign to the
charge anomaly, $Trace(Q_A) \propto \xi_A $.  A superpotential
monomial, $ W _x= \prod _ {i=1}^n \phi _i^ {s_i ^x } $, is allowed if
it carries vanishing anomaly free and anomalous charges, $\sum
_{i=1}^n Q_{a} (\phi _i ) s_i^x =0, \ \sum _{i=1}^n Q_{A} (\phi _i )
s_i ^x =0 $, and satisfies the string selection rules listed in
Eq.~(\ref{eqsel}).
% of Appendix~\ref{appexb}.   

We define the order $n$ of a holomorphic D flat or superpotential
monomial, $ P _\a (\hat \phi _i ) $ or $ W_x (\hat \phi _k , \phi _l)
$, by the number of field factors this contains, independently of the
values of the fields positive integer exponents, $r_i ^\a ,\ s_i ^x $.
The number of inequivalent monomial solutions rises fast with the
order $n$ considered, reaching orders of hundreds already at the
relatively low order, $n = 4$.  The solutions will be ordered
according to increasing values of the number of field factors, $ n
=1,\ 2,\ \cdots $, while allowing the power exponents to range over
the discrete set of values, $ r_i ^\a , \ s_i ^x \in [0,\ 1,\ 2 ]$.
Note that with this definition, the order $n$ of a monomial is always
bounded by its effective order, $\sum _{i=1}^n r_i ^\a $ or $\sum
_{i=1}^n s_i^x $.  To avoid dealing with an overwhelming number of
solutions, we generally restrict to orders $n \leq 4$.

The unbroken (anomaly free) Abelian symmetries $ U(1)_x$ along a flat
direction are identified by the kernel of the charge matrix, $ A_{ai}
= Q _a ( \hat \phi _i) , \ [a=1, \ N_a -1 ] $ where $\hat \phi _i$
ranges over the modes excited along the flat direction and the index
label $ a $ ranges over the anomaly free Abelian symmetries. In
practice, we proceed by evaluating the left eigenvectors with zero
eigenvalues of the matrix, $ A_{ai} , \ [a, i= 1, \cdots , N_a -1 ]$
namely, $x ^T A = 0$, for some selected subset of the $\hat \phi _i$,
and next retain among the Abelian charges, $ Q_{x} = \sum _a x_a Q_a
$, those yielding zero charges, $ Q_x (\hat \phi _i) =0 $, for the
full set of $\hat \phi _i $.

The effective trilinear interactions descending from
non-renormalizable operators of order $n +3$ are assigned string tree
level coupling constants given by the approximate
formula~\cite{cleav398} \be W _3^{(n)} = \tilde c^{(n)}_{ijk} (\phi )
\Phi _i \Phi _j \Phi _k ,\ \left[\tilde c^{(n)}_{ijk} (\phi ) =
c^{(n)}_{ijk} < \prod _{i=1}^n \phi _i > , \ c^{(n)}_{ijk} \simeq g_s
{C_n I_n \over (\pi M_\star )^n } \right]
\label{eqestc}  
\ee where $ C_n $ are constant coefficients of $ O(1)$ and $ I_n =
\int \prod _{i=1} ^nd^2 z_i f (z_i, \bar z_i ) $ are integrals over
the location of vertex operators on the world sheet sphere
surface. The integrals for the four and five point string amplitudes
evaluate numerically to~\cite{kalara91,faraggi97,cvet98}: $ I_1\simeq
70, \ I_2 \simeq 400 $.  The string coupling constant is commonly set
at, $ g_s = g_X /\sqrt 2 \simeq 1/2.$ For the case of a 0-d flat
direction, assigning the singlet fields the common VEV $\hat \phi $
determined by the Fayet-Iliopoulos term, one infers the following
formula for the coefficients \be \vert \hat \phi \vert \simeq \left(-
{\xi _A \over Q_A (\phi ) } \right)^\ud \simeq {g_s M_\star \over \pi
} \left( -{Trace(Q_A) \over 192 Q_A ( \hat \phi ) \sqrt {k_A} }
\right) ^\ud \ \Longrightarrow \ \tilde c^{(n)}_{ijk} \simeq {g_s
^{n+1} C_n I_n \over (\pi ^2 \sqrt {192} )^n } \left( -{Trace(Q_A)
\over Q_A ( \hat \phi ) \sqrt {k_A} } \right) ^ {n /2} .  \ee With
increasing order of the non-renormalizable operators, we may thus
anticipate a strong or mild suppression depending on the unknown
extrapolation of $ I_n$ for large $n$ and the model dependent size of
the anomalous charges.  Using, say, $ I_n = 10^3, \ C_n =1$, and
dropping the factor depending on $Q_A$ by setting, $ { -Trace ( Q_A )
\over \sqrt {k_A} Q_A (\hat \phi ) } \to 1$, gives, $\tilde
c^{(n)}_{ijk} \simeq 10^{-2n } \ g_s ^{1+n} C_n I_n \approx 10^{-2n
+3} /2 ^{n+1} .$ A similar formula holds for the $\mu $-term
coefficient of the electroweak Higgs bosons bilinear coupling, $\tilde
c^{(n)} _{ij} h_i h_j $, by substituting, $\tilde c^{(n)} _{ij} = \mu
^{(n)} / M_\star $.  Bounding the coupling constant by the requisite
ratio between Planck and effective supersymmetry breaking mass scales,
$\mu ^{(n)} / M_\star \leq \text{TeV} /M_\star \simeq 10^{-15} $,
yields, $n \geq 8$, which shows that non-renormalizable effective
operators starting from $ O(\hat \phi ^ {(8)} ) h_i h_j $ can be
tolerated.  Considering instead the tentative extrapolation, $ I_n
\simeq 10^{n\over 2} , \ [C_n =1]$, and the numerical estimates for
the anomalous charge appropriate to model $A1$ (to be detailed below)
\be \left( { -Trace ( Q_A ) \over \sqrt {k_A} Q_A (\hat \phi ) }
\right) ^\ud \simeq 10, \ \ {\vert \hat \phi \vert \over g_s M_\star }
\simeq {1\over \pi \sqrt {192} } \left( { -Trace ( Q_A ) \over \sqrt
{k_A} Q_A (\hat \phi ) } \right) ^\ud \simeq 0.23 , \ee then the
corresponding condition on the $\mu $-term coupling, $\mu ^{(n)} /
M_\star \simeq \ud 10^{-n} \leq 10 ^{-15} $, gives the stronger bound
on the order of non-renormalizable operators, $n \geq 14$.  Note that
lowering $\hat \phi _i $ should weaken the bounds on $n$.

\section{Singlet extended   supersymmetric standard models}
\label{sec2}

The Pati-Salam models $A1, \ A2,\ B$ of~\cite{Kobayashi:2004ya} for
the orbifold $ Z_{6-II} $ of $ T^6 $ with two Wilson lines only (since
$W'_2=0$), are described by the 16-d shift vectors for the gauge twist
and Wilson lines of the $E_8 \times E_8$ group lattice: \bea &&
\text{Model} \ A1: \ V= {1\over 6} ( 2, 2, 2, 0^5 ) (1, 1, 0^6) , \
W_2= {1\over 6} (3, 0^4, 3^3) (0^8) , \ W_3= {1\over 6} (2, -2, 0^6 )
( 0^2, 4, 0^5) .  \cr && \text{Model} \ A2: \ V= {1\over 6} ( 2^3, 0^5
) (1^2 , 0^6 ) , \ W_2= {1\over 6} ( 3, 0^4 , 3^3 ) (0^8 ) , \ W_3=
{1\over 6} ( 4, 2, -2, 0 ^5) (0, 4, 2^2 , 0^4 ) . \cr && \text{Model}
\ B: \ V= {1\over 6} (4, 1^2 , 0^5 ) (2^2, 0^6 ) , \ W_2= {1\over 6}
(0^3 , 3^2 , 0^3 ) (3, 0, 3, 0^5 ) , \ W_3= {1\over 6} ( 0, 2, -2, 0^5
) ( 0^2 , 4, 0^5 ) . \eea The 4-d gauge groups for the above three
models are: \bea \text{Model} \ A_1 &:& SU(4)\times SU(2)_1\times
SU(2)_2 \times U(1) ^{5} \times (SO(10)' \times SU(2)') . \\
\text{Model} \ A_2 &:& SU(4)\times SU(2)_1\times SU(2)_2 \times U(1)^
5 \times (SO(10)'\times SU(2) ' ) . \\ \text{Model} \ B &:&
SU(4)\times SU(2) _1\times SU(2)_2 \times U(1)^6 \times SO(10)'.  \eea

%The   massless  string  spectra   for  models $ A1, \ A2$  and $ B $ 
%are displayed in Tables~\ref{tabmoda1},~\ref{tabmdta2}
%and~\ref{tabmodeleb}, respectively.  

Our results for  the  massless  string  spectra   of  models 
$ A1, \ A2$  and $ B $   match  to  those  of~\cite{Kobayashi:2004ya},  except   for 
certain assignments of  oscillator numbers, $ N_{I, \bar I} ^L  $,
hence of  $R_I$ charges.      The  numerical  data  for the    linear   and cubic traces of  the
anomalous $ U(1)_A$    gauge charge   and the  Fayet-Iliopoulos term,
$  \xi _A  /(g_s M_\star ) ^2 \equiv   Trace (Q_A) /(192 \pi ^2 \sqrt
{ k_A} ) $  are  displayed in the following table. This also
includes results for  the  total number of massless  string states
$\Phi _i$  and the   degeneracies of the various   modes. 
\vskip 0.5 cm \begin{center} \begin{tabular}{|c|ccccc||ccccccccc|} 
%\hline
Model & $  Tr (Q_A ) $  & $    Tr (Q_A^3 )  $  & $  

Tr (Q_A Q_a ^2 ) /2 $  & $    k_A   $  & $     \xi _A  /(g_s M_\star
  ) ^2  $   & $ \  \Phi _i \  $  &  $ \   (S_i, Y_i)  \  $ &   $ \
h_i \  $   &  $  \  C_i  \  $ & $  \  f_i  \  $ & $ \   \bar f_i  \  $ 
& $  \   f^c_i \   $ & $ \  \bar f^c_i \  $ &  $ \   (q_i, \bar q_i) \
  $ \\ \hline 
$A1 $  & $     -2 $  & $      -0.00437 $  & $ -0.16667  $  & $
0.01750   $  & $      - 0.8 \  \times  10^{-2}$  & 78 
&  34 &  7 & 9  & 3  &0   & 5  &2   &4 \\ 
$A2$  & $    -131.4 $  & $  -393.90 $  & $    -10.94  $ & $    24  $
& $  - 1.4 \ \times 10^{-2} $  & 77 
&  36 & 3  & 3  & 3 &0  &4  & 1  & 6   \\     
$B $  & $    13.66 $  & $   0.312 $  & $   1.139  $  & $    0.182  $
& $ 1.7\ \times 10^{-2} $  & 84 &  48  & 6 &3  &6 &3  &6  &3  & 4  \\   
%\hline
\end{tabular} \end{center}   \vskip 0.5 cm

\subsection{Model $A1$} 
\squeezetable \begin{table}[H] \begin{center}
\caption{ \label{tabmoda1} Spectrum of string massless modes for the
Pati-Salam model $A1$ of~\cite{Kobayashi:2004ya} with gauge group, $
G_{422} \times (SO(10)'\times SU(2) ') \times U(1)^5$. The pair of
group factors $SU(2)_{1,2} $ identify with $ SU(2)_{L,R} $.  The
non-Abelian singlet modes $\phi _i \sim (1,1,1) (1,1)' $ without and
with oscillator excitations are denoted by $ S_i $ and $Y_i $.  The
non-singlet modes include the doublets, $ d^l_i \sim (1,2,1)(1,1)',\
d^r _i \sim (1,1,2) (1,1)',\ d' _i \sim (1,1,1) (1',2')$; the
bidoublets, $ h _i \sim (1,2,2)(1,1)'$;
%h ^{'l}_i   \sim (1,2,1) (1',2'),\ h ^{'r}_i   \sim (1,1,2) (1',2')
   the quartets, $ q_i \sim (4,1,1)(1,1)', \ \bar q_i \sim (\bar
4,1,1)(1,1)' $; the sextets, $ C_i \sim (6,1,1)(1,1)' $; the
bifundamentals, $ f^c_i \sim (\bar 4,1,2)(1,1)',\ \bar f^c_i \sim
(4,1,2)(1,1)',\ f _i \sim (4,2,1)(1,1)',\ \bar f _i \sim (\bar
4,2,1)(1,1)',\ $; the decuplets, $ T'_i \sim (1,1,1) (10',1')$; and
the sextuplets, $ \S '_i \sim (1,1,1) (16',1') ,\ \bar \S '_i \sim
(1,1,1) (\overline{16'},1')$. The first column lists the twisted
sectors label $g$; the second column lists the modes name; the third
column consists of three subcolumns which list the twisted subsector
specified by the excited Wilson lines, with $ N_{32} = 1 , \ [ g=0] ;
\ \ N_{32} = 3n_2 + n_3 +1, \ [ g=1] ; \ \ N_{32} = n_3+1 , \ [ g=2,4]
$; and $ N_{32} = n_2 +1, \ [ g=3] $, the complex phase $\g (g)$
describing the orbifold twist action on the fixed points in the $
T^2_1$ plane and the modes multiplicity (degeneracy) $\cald $; the
third column consists of three subcolumns which list the shifted
$H$-momenta specified by the charges, $ 6 R^I , \ [ I=1,2,3]$; and the
fourth column consists of five subcolumns which list the Abelian gauge
charges in the rotated basis starting with the anomaly free charges
$Q_a $ and ending with the single anomalous charge $Q_A$.  In order to
quote integer charges, the Abelian charges have been rescaled as,
% $ Q_a (\phi _i ) \to Q_a (\phi _i ) /min ( Q_a (\phi _k ) ) $.  
 $ Q_1/0.1667, \ Q_2/0.02083 \ , \ Q_3/0.0104 , \ 4\times Q_4/0.006944
, \ Q_A/0.008333. $ (Note that the first column entry for $g$ is
omitted in the right-hand part of
the table.)}
\vskip 0.5 cm  
\begin{tabular}{|ccccccccccccccc||cccccccccccccc|}\hline $g$ & M  & $ N_{32}$ & $\gamma
$ & $\cald$ && $ 6 (R_1$ & $R_2$ & $R_3 )$ && $Q_1$ & $Q_2$ & $Q_3$ & $Q_4$ & $Q_A$ & M & $ N_{32}$ & $\gamma$ & $\cald$ && $6 (R_1$ & $R_2$ & $R_3)$ && $Q_1$ & $Q_2$ & $Q_3$ & $Q_4$ & $Q_A$ \\   \hline 
  0 & $h_ { 1}$ & $1$ & $1$ & $1$ && $0$ & $6$ & $0$ && $0$ & $3$ & $-3$ & $-6$ & $3$ 
& $S_ { 1}$ & $1$ & $1$ & $1$ && $0$ & $6$ & $0$ && $0$ & $0$ & $-6$ & $24$ & $-12$ \\ 
& $T'_{ 1}$ & $1$ & $1$ & $1$ && $6$ & $0$ & $0$ && $0$ & $0$ & $-3$ & $12$ & $-6$
& $\bar f^c_{ 1}$ & $1$ & $1$ & $1$ && $0$ & $0$ & $6$ && $3$ & $3$ & $0$ & $0$ & $0$ \\ 
& $f_ { 1}$ & $1$ & $1$ & $1$ && $6$ & $0$ & $0$ && $3$ & $0$ & $3$ & $6$ & $-3$ 
& $f^c_{ 1}$ & $1$ & $1$ & $1$ && $0$ & $0$ & $6$ && $-3$ & $-3$ & $0$ & $0$ & $0$ \\ 
 \hline 
  1 & $Y_ { 1}$ & $1$ & $1$ & $2$ && $1$ & $2$ & $9$ && $2$ & $2$ & $-1$ & $4$ & $-2$
& $Y_ { 2}$ & $1$ & $1$ & $2$ && $1$ & $2$ & $-3$ && $2$ & $2$ & $-1$ & $4$ & $-2$ \\ 
& $Y_ { 3}$ & $1$ & $1$ & $2$ && $-5$ & $-4$ & $3$ && $2$ & $2$ & $-1$ & $4$ & $-2$
& $Y_ { 4}$ & $1$ & $1$ & $2$ && $-17$ & $2$ & $3$ && $2$ & $2$ & $-1$ & $4$ & $-2$ \\ 
& $Y_ { 5}$ & $1$ & $1$ & $2$ && $-5$ & $2$ & $3$ && $2$ & $-4$ & $-1$ & $4$ & $-2$
& $Y_ { 6}$ & $1$ & $1$ & $2$ && $-5$ & $2$ & $3$ && $-4$ & $-1$ & $2$ & $10$ & $-5$ \\ 
& $Y_ { 7}$ & $1$ & $1$ & $2$ && $-5$ & $2$ & $3$ && $-4$ & $-1$ & $-4$ & $-2$ & $1$
& $f^c_{ 2}$ & $1$ & $1$ & $2$ && $1$ & $2$ & $3$ && $-1$ & $-1$ & $-1$ & $4$ & $-2$ \\ 
& $f_ { 2}$ & $1$ & $1$ & $2$ && $1$ & $2$ & $3$ && $-1$ & $-1$ & $-1$ & $4$ & $-2$
& $Y_ { 8}$ & $2$ & $1$ & $2$ && $-5$ & $2$ & $3$ && $4$ & $1$ & $-2$ & $-10$ & $-5$ \\ 
& $d'_{ 1}$ & $2$ & $1$ & $2$ && $1$ & $2$ & $3$ && $4$ & $1$ & $1$ & $-4$ & $7$
& $Y_ { 9}$ & $2$ & $1$ & $2$ && $-5$ & $2$ & $3$ && $-2$ & $-2$ & $1$ & $-4$ & $-8$ \\ 
& $d'_{ 2}$ & $2$ & $1$ & $2$ && $1$ & $2$ & $3$ && $-2$ & $-2$ & $4$ & $2$ & $4$
& $Y_ {10}$ & $3$ & $1$ & $2$ && $-5$ & $2$ & $3$ && $0$ & $3$ & $0$ & $18$ & $1$ \\ 
& $d'_{ 3}$ & $3$ & $1$ & $2$ && $1$ & $2$ & $3$ && $0$ & $3$ & $3$ & $-12$ & $1$
& $Y_ {11}$ & $3$ & $1$ & $2$ && $-5$ & $2$ & $3$ && $0$ & $-3$ & $0$ & $18$ & $1$ \\ 
& $d'_{ 4}$ & $3$ & $1$ & $2$ && $1$ & $2$ & $3$ && $0$ & $-3$ & $3$ & $-12$ & $1$
& $\bar q_{ 1}$ & $4$ & $1$ & $2$ && $-5$ & $2$ & $3$ && $-1$ & $2$ & $-1$ & $4$ & $-2$ \\ 
& $q_ { 1}$ & $4$ & $1$ & $2$ && $1$ & $2$ & $3$ && $-1$ & $-1$ & $2$ & $10$ & $-5$ 
& $q_ { 2}$ & $4$ & $1$ & $2$ && $1$ & $2$ & $3$ && $-1$ & $-1$ & $-4$ & $-2$ & $1$ \\ 
& $d^l_{ 1}$ & $4$ & $1$ & $2$ && $1$ & $2$ & $3$ && $2$ & $2$ & $2$ & $10$ & $-5$
& $d^l_{ 2}$ & $4$ & $1$ & $2$ && $1$ & $2$ & $3$ && $2$ & $2$ & $-4$ & $-2$ & $1$ \\ 
& $d^r_{ 1}$ & $4$ & $1$ & $2$ && $1$ & $-4$ & $3$ && $2$ & $-1$ & $-1$ & $4$ & $-2$
& $d^r_{ 2}$ & $4$ & $1$ & $2$ && $-11$ & $2$ & $3$ && $2$ & $-1$ & $-1$ & $4$ & $-2$ \\ 
& $d^l_{ 3}$ & $4$ & $1$ & $2$ && $-5$ & $2$ & $3$ && $-4$ & $-1$ & $-1$ & $4$ & $-2$
& $\bar q_{ 2}$ & $5$ & $1$ & $2$ && $1$ & $2$ & $3$ && $1$ & $1$ & $-2$ & $-10$ & $-5$ \\ 
& $d^r_{ 3}$ & $5$ & $1$ & $2$ && $1$ & $2$ & $3$ && $-2$ & $1$ & $1$ & $-4$ & $-8$
& $d^l_{ 4}$ & $5$ & $1$ & $2$ && $1$ & $2$ & $3$ && $-2$ & $-2$ & $-2$ & $-10$ & $-5$ \\ 
& $d^r_{ 4}$ & $6$ & $1$ & $2$ && $-5$ & $2$ & $3$ && $0$ & $0$ & $0$ & $18$ & $1$
& $h ^{'r}_{ 1}$ & $6$ & $1$ & $2$ && $1$ & $2$ & $3$ && $0$ & $0$ & $3$ & $-12$ & $1$ \\ 
 \hline 
  2 & $S_ { 2}$ & $1$ & $-1$ & $1$ && $2$ & $4$ & $0$ && $4$ & $4$ & $-2$ & $8$ & $-4$
& $Y_ {12}$ & $1$ & $1$ & $2$ && $-4$ & $4$ & $0$ && $4$ & $-2$ & $-2$ & $8$ & $-4$ \\ 
& $Y_ {13}$ & $1$ & $-1$ & $1$ && $2$ & $10$ & $0$ && $4$ & $-2$ & $-2$ & $8$ & $-4$
& $S_ { 3}$ & $1$ & $-1$ & $1$ && $2$ & $4$ & $0$ && $4$ & $-2$ & $4$ & $-16$ & $8$ \\ 
& $C_ { 1}$ & $1$ & $-1$ & $1$ && $2$ & $4$ & $0$ && $-2$ & $-2$ & $-2$ & $8$ & $-4$
& $f^c_{ 3}$ & $1$ & $1$ & $2$ && $2$ & $4$ & $0$ && $1$ & $1$ & $-2$ & $8$ & $-4$ \\ 
& $Y_ {14}$ & $2$ & $1$ & $2$ && $-4$ & $4$ & $0$ && $0$ & $0$ & $0$ & $0$ & $-10$
& $Y_ {15}$ & $2$ & $-1$ & $1$ && $2$ & $10$ & $0$ && $0$ & $0$ & $0$ & $0$ & $-10$ \\ 
& $T'_{ 1}$ & $2$ & $-1$ & $1$ && $2$ & $4$ & $0$ && $0$ & $0$ & $0$ & $18$ & $-4$
& $d'_{ 5}$ & $2$ & $1$ & $2$ && $2$ & $10$ & $0$ && $0$ & $0$ & $3$ & $6$ & $2$ \\ 
& $d'_{ 6}$ & $2$ & $-1$ & $1$ && $-4$ & $4$ & $0$ && $0$ & $0$ & $3$ & $6$ & $2$
& $S_ { 4}$ & $2$ & $-1$ & $1$ && $2$ & $4$ & $0$ && $0$ & $0$ & $6$ & $-24$ & $2$ \\ 
& $\S '_{ 1}$ & $2$ & $1$ & $2$ && $2$ & $4$ & $0$ && $0$ & $0$ & $3$ & $-3$ & $-1$
& $S_ { 5}$ & $2$ & $-1$ & $1$ && $2$ & $4$ & $0$ && $0$ & $0$ & $-6$ & $-12$ & $-4$ \\ 
& $d'_{ 7}$ & $2$ & $1$ & $2$ && $2$ & $4$ & $0$ && $0$ & $0$ & $-3$ & $-6$ & $8$
& $S_ { 6}$ & $3$ & $-1$ & $1$ && $2$ & $4$ & $0$ && $-4$ & $2$ & $-4$ & $16$ & $2$ \\ 
& $d'_{ 8}$ & $3$ & $1$ & $2$ && $2$ & $4$ & $0$ && $-4$ & $2$ & $-1$ & $-14$ & $2$
&&&&&&&&&&&&&&   \\ \hline 
  3 & $S_ { 7}$ & $1$ & $\o$ & $2$ && $3$ & $0$ & $3$ && $0$ & $3$ & $-6$ & $6$ & $-3$
& $S_ { 8}$ & $1$ & $\o ^2$ & $2$ && $3$ & $0$ & $3$ && $0$ & $3$ & $0$ & $-18$ & $9$ \\ 
& $h_ { 2}$ & $1$ & $\o ^2$ & $2$ && $3$ & $0$ & $3$ && $0$ & $0$ & $-3$ & $12$ & $-6$
& $h_ { 3}$ & $1$ & $1$ & $4$ && $3$ & $0$ & $3$ && $0$ & $0$ & $3$ & $-12$ & $6$ \\ 
& $C_ { 2}$ & $1$ & $\o ^2$ & $2$ && $3$ & $0$ & $3$ && $0$ & $0$ & $-3$ & $12$ & $-6$
& $C_ { 3}$ & $1$ & $1$ & $4$ && $3$ & $0$ & $3$ && $0$ & $0$ & $3$ & $-12$ & $6$ \\ 
& $S_ { 9}$ & $1$ & $1$ & $4$ && $3$ & $0$ & $3$ && $0$ & $-3$ & $0$ & $18$ & $-9$
& $S_ {10}$ & $1$ & $\o$ & $2$ && $3$ & $0$ & $3$ && $0$ & $-3$ & $6$ & $-6$ & $3$ \\ 
 \hline 
  4 & $C_ { 4}$ & $1$ & $1$ & $2$ && $4$ & $2$ & $0$ && $2$ & $2$ & $2$ & $-8$ & $4$
& $S_ {11}$ & $1$ & $1$ & $2$ && $4$ & $2$ & $0$ && $-4$ & $2$ & $-4$ & $16$ & $-8$ \\ 
& $Y_ {16}$ & $1$ & $1$ & $2$ && $4$ & $-4$ & $0$ && $-4$ & $2$ & $2$ & $-8$ & $4$
& $Y_ {17}$ & $1$ & $-1$ & $1$ && $10$ & $2$ & $0$ && $-4$ & $2$ & $2$ & $-8$ & $4$ \\ 
& $S_ {12}$ & $1$ & $1$ & $2$ && $4$ & $2$ & $0$ && $-4$ & $-4$ & $2$ & $-8$ & $4$
& $\bar f^c_{ 2}$ & $1$ & $-1$ & $1$ && $4$ & $2$ & $0$ && $-1$ & $-1$ & $2$ & $-8$ & $4$ \\ 
& $d'_{ 9}$ & $2$ & $-1$ & $1$ && $4$ & $2$ & $0$ && $4$ & $-2$ & $1$ & $14$ & $-2$
& $S_ {13}$ & $2$ & $1$ & $2$ && $4$ & $2$ & $0$ && $4$ & $-2$ & $4$ & $-16$ & $-2$ \\ 
& $d'_{10}$ & $3$ & $-1$ & $1$ && $4$ & $2$ & $0$ && $0$ & $0$ & $3$ & $6$ & $-8$
& $S_ {14}$ & $3$ & $1$ & $2$ && $4$ & $2$ & $0$ && $0$ & $0$ & $6$ & $12$ & $4$ \\ 
& $S_ {15}$ & $3$ & $1$ & $2$ && $4$ & $2$ & $0$ && $0$ & $0$ & $-6$ & $24$ & $-2$
& $d'_{11}$ & $3$ & $1$ & $2$ && $10$ & $2$ & $0$ && $0$ & $0$ & $-3$ & $-6$ & $-2$ \\ 
& $d'_{12}$ & $3$ & $-1$ & $1$ && $4$ & $-4$ & $0$ && $0$ & $0$ & $-3$ & $-6$ & $-2$
& $T'_{ 2}$ & $3$ & $1$ & $2$ && $4$ & $2$ & $0$ && $0$ & $0$ & $0$ & $-18$ & $4$ \\ 
& $Y_ {18}$ & $3$ & $1$ & $2$ && $4$ & $-4$ & $0$ && $0$ & $0$ & $0$ & $0$ & $10$
& $Y_ {19}$ & $3$ & $-1$ & $1$ && $10$ & $2$ & $0$ && $0$ & $0$ & $0$ & $0$ & $10$ \\ 
& $\bar\S '_{ 1}$ & $3$ & $-1$ & $1$ && $4$ & $2$ & $0$ && $0$ & $0$ & $-3$ & $3$ & $1$
&&&&&&&&&&&&&& \\ \hline \hline
\end{tabular} 
  \end{center}  \end{table}

The massless string spectrum of model $ A1 $ is 
displayed in Table~\ref{tabmoda1}.  The set of holomorphic
invariant monomials of order $ n\leq 2 $ consist of eleven solutions,
$ P_\a =[Y_ {11} S_ { 8} ,\ S_ { 3} S_ { 6} ,\ Y_ {14} Y_ {18}^2 ,\ Y_
{14}Y_ {19}^2 ,\ Y_ {15} Y_ {19}^2 ,\ Y_ {15} Y_ {18}^2,\ Y_ {18}^2 ,\
Y_ {18} Y_ {19} ,\ Y_ {18}^2Y_ {19}, \ Y_ {18}Y_ {19}^2, \ Y_ {19}^2
]$.  None of these D flat directions is found to be lifted by the
superpotential couplings in $ W_s$ at orders $n \leq 4$.  Rather than
analyzing each monomial $P_\a $ individually, we combine these
multiplicatively into what we term as the `composite' D-flat
direction, $ \calp ^{(2)} = P_1 \cdots P_{11}$, which then includes
the eight modes, $ Y_ {11}, \ S_ { 8}, \ S_ { 3}, \ S_ { 6}, \ Y_
{14}, \ Y_ {18}, \ Y_ {19}, \ Y_ {15} $.  The set of excluded
couplings along the composite direction is clearly the union of all
the sets excluded by the individual couplings of order $n$.  All these
directions are $0 $-d ones, hence with a common VEV given by, $\vert
\hat \phi \vert / (g_s M_\star ) \simeq 0.23 $.  The effective
superpotential components obtained by assigning all the fields along
$\calp ^{(2)} $ a common value $\hat \phi _i = \phi $ read at order
$n\leq 4$ \bea W_s^{(2)} &=& \phi S_ { 1} S_ {10} +\phi ^3 Y_ { 3} Y_
{ 9} +\phi ^2 S_ {11} S_ {13} +\phi ^2 S_ {10} S_ {15} .  \cr && \cr W
_h &=& h_1 h_1 [\phi^ 5Y_ {17} +\phi^ 5S_ { 4} ] + h_2 h_2 [ \phi^ 4S_
{10}] + h_3 h_3 [S_ { 1} ] + h_1h_2 [\phi^ 2S_ {10} ] + h_1 h_3 [\phi^
4 + S_ { 9} ] + h_2 h_3 [\phi^ 6 +\phi^ 4S_ { 9} ] . \cr && \cr W_C
&=& C_1 C_1 [\phi^ 3Y_ {10} ] +C_2 C_2 [\phi^ 4S_ {10} ] +C_3 C_3 [S_
{ 1} ] +C_4 C_4 [\phi^ 4Y_ { 7} ] +C_1 C_2 [ 0 ] +C_1 C_3 [\phi^ 2Y_ {
3} ] +C_1 C_4[\phi Y_ { 3}Y_ { 9}] \cr && +C_2 C_3 [\phi^6+\phi^ 4S_ {
9} ] +C_2 C_4 [\phi^ 5Y_ { 9} ] +C_3 C_4 [ Y_ {5}S_ {11} ] . \cr &&
\cr W _q &=& q_1 \bar q_1 [ 0] + q_1 \bar q_2 [\phi^ 5 ] + q_2 \bar
q_1 [ 0] + q_2 \bar q_2 [ S_ {14} ] .  \cr && \cr W_f &=& f^c_1 \bar
f^c_1 [ 0 ] + f^c_1 \bar f^c_2 [ S_ { 2} ] + f^c_2 \bar f^c_1[ 0 ] +
f^c_2 \bar f^c_2[\phi ^ 2Y_ { 3}] + f^c_3 \bar f^c_1 [ S_ {12} ] +
f^c_3 \bar f^c_2 [ 0 ] .  \cr && \cr W_{hff^c} &=& h_1 f_1 f^c_1 [ 1 ]
+ h_1 f_1 f^c_2 [ \phi Y_ { 9} ]+ h_1 f_1 f^c_3 [ \phi^ 2Y_ { 9} ] +
h_1 f_2 f^c_1 [0 ] + h_1 f_2 f^c_2 [ 0 ] + h_1 f_2 f^c_3 [ 0 ] \cr &&
+ h_2 f_1 f^c_1 [ \phi^ 4 ] + h_2 f_1 f^c_2 [\phi ^ 3Y_ { 9} ] + h_2
f_1 f^c_3 [ \phi^ 4Y_ { 9} ] + h_2 f_2 f^c_1 [ 0] + h_2 f_2 f^c_2 [0]+
h_2 f_2 f^c_3 [ 0 ] \cr && + h_3 f_1 f^c_1 [ 0 ] + h_3 f_1 f^c_2 [ 0 ]
+ h_3 f_1 f^c_3 [Y_ { 7}] + h_3 f_2 f^c_2 [Y_ { 1} ] + h_3 f_2 f^c_1
[Y_ { 1} + Y_ { 2} + Y_ { 4}] + h_3 f_2 f^c_3 [\phi^ 2 ] .
\label{eqsup5} 
\eea 
    The results  for $ W_h $  through $ W_{hff^c} $  are 
illustrative ones in the sense that they include  only a subset of
the couplings at most linear  or quadratic in the singlets  which are
not part of the flat direction. For convenience, we have listed the  
results for the  bilinear couplings  along $\calp ^{(2)} $ in the second
column of Table~\ref{dirplatesa1}.   The flavor bases for the massless
fermions consist of the 3 modes $ f_{1,2}$ and the 3 linear
combinations of the 5 modes $ f^c_{1,2,3} $ which are left unpaired.
The fermion mass matrices are sums of $ 3\times 5$ matrices $ <h_k>
f_{1,2} f^c_{1,2,3} , \ [k=1,2,3 ]$ involving the combinations of 7
light bidoublets which acquire finite VEVs at the electroweak
transition.  The presence of an unsuppressed coupling $ h_1 \ f_1 f^c
_1$ suggests that $ h_1$ should be the dominant component of the light
electroweak doublets and $ f_1,\ f^c _1$ should be associated to the
third generation fermions.  The degenerate pairs of modes $ f_2 $ and
$ f^c _2$ must then be associated to the first two generations.  As
long as the Dihedral $D_4$ discrete symmetry
of~\cite{Kobayashi:2004ya,ko07} is unbroken, the fermions mass
matrices $ f_2 f_i ^c $ would include pairs of identical rows, which
means that their rank is $ \leq 2$.  This seemingly good starting
point motivates one to search for mechanisms breaking the $D_4$
symmetry.  As discussed in~\cite{Kobayashi:2004ya,ko07}, the
non-renormalizable operators, $ h_k f_i f^c_j \prod _{k, l, m, m'} S_k
Y_l {\cal O}^c _{m m'} $, involving the composite singlet fields, $
{\cal O}^c _{ij} = f^c _i\bar f^c _j $, produce contributions at the
$PS\to SM $ transition which can improve predictions for the fermions
mass matrices.  We note that an alternative way to break the $D_4$
symmetry is by switching on the third Wilson line, $ n'_2W'_2$.
 
That no constant or linear terms appears in $W_s^{(2)}$ indicates that
all the $n=2$ flat directions are indeed unlifted.  The decoupled
massive singlets are, $S_ { 1, 10, 11, 13,15} $ and $ Y_ {3, 9} $.
The $\mu $-term bilinear coupling in $W_h$ is a rank $2$ matrix with
the massless mode, $ ( h_2 - \phi ^2 h_1)$.  However, the suppressed
component is along $ h_1$, which is the preferred mode whose VEV
supposedly dominates the fermion masses.  The trilinear couplings $ W=
(\phi ^4 h_2h_2 + \phi ^2 h_1h_2 ) S _{10} $ are seen to select the
massive singlet $S_{10}$, which figures among the modes which acquire
large masses along $\calp ^{(2)}$. However, it should be noted that
$S_{10}$ remains massless for a subset of the $n=2$ directions $P_\a
$.  The above result for $ W_f$ shows that no mass pairings among the
matter modes are allowed.  The mass matrices $ M^f _{ij} f_i f^c _j $
have the single non-vanishing entry, $ M^f _{11} = <h_1>,$ if we omit
the massive $h_3$ mode.  Examination of the exotic modes mass
couplings, shows that $ q_1, \ \bar q_2$ and $C_2, \ C_3$ pair up,
while the pairs $ q_2, \ \bar q_1$ and $ C_1, \ C_4$ remain massless.
We also find that except for the coupling, $ C_1 \bar f^c_2\bar f^c_2
[\phi^ 2 ]$, all other trilinear couplings, $ C_1 f_i^c f_j^c ,\ C_1
\bar f_i^c \bar f_j^c $ in $W_{Cff} $ are suppressed.

The $ 4 \times 8 $ matrix of the Abelian charges for modes along $
\calp ^{(2)}$ is found to have column rank $2$.  The existence of a
secluded sector requires finding three singlets, $\phi _I, \ \phi _J,
\ \phi _K,$ charged under the unbroken $ U(1)_x$ with a mildly
suppressed trilinear coupling, $c(\hat \phi _k) \phi _I \phi _J \phi
_K $.  A wide range of choices is clearly possible given the sizeable
number of singlets orthogonal to the moduli space of $ \calp ^{(2)} $,
and the abundant number of monomial solutions in $W_s$.  An inspection
of the allowed couplings lets us select the following effective cubic
or quartic order couplings $ Y_ { 1} Y_ { 2} S_ {12} , \ Y_ { 8} S_ {
7}S_ {12} <S_ {14}> ,\ <Y_ { 4}^2 > Y_ {12} Y_ {16} S_ {12},\ Y_ {10}
S_ { 7} S_ {12} <S_ {13} > $, that can give rise to a secluded sector
with three massless singlets coupled by mildly suppressed trilinear
interactions.  A systematic search does not appear warranted at this
stage.

The special role of $S_{10} $ motivates us to examine its low order
self-couplings, which are given for $ n \leq 4 $ by $ W_s (S_{10}) =
S_{10} [0] + S_{10} ^2 [\phi^ 3S_ { 7} ] + S_{10} ^3 [\phi^ 4S_ { 7}
^2 ] .$ These resuls indicate that a cubic self-coupling, $ S_{10} ^3
\phi^ 4 <S_ { 7} ^2 > $, cannot arise without an unacceptably large
mass term, $S_{10} ^2 \phi^ 3 <S_ { 7} > $.  Since this correlation
between the quadratic and cubic couplings of $ S_{10} $ also holds for
the individual flat directions, we conclude that the occurrence of a
trilinear coupling, $ h_i h_j S_{10}$, with a finite cubic
self-coupling is not the favored option.

We also found useful to study a class of solutions on an individual
basis.  We present in Table~\ref{dirplatesa1} a subset of the bilinear
couplings for four randomly selected F flat monomial solutions of
order $n=4$.  The monomials $ P^{(4)} _{ II} ,\ P^{(4)} _{ III},\
P^{(4)} _{ IV} $ are seen to allow a pair of massless bidoublets, $
h_1, \ h_2$, with preferred trilinear coupling, $ W_{h} = h_1 h_2
S_{10}$.  Along $P^{(4)} _{ II}$, the diagonal trilinear coupling, $
W_{h} =\phi ^5 h_2 h_2 S_{13} $, is also present.  The direction $
P^{(3)}_{I} $ has all three bidoublets massless.  However, some
subsets of the exotic modes always remain massless.

%%%%%%%%%%%%%%%%%%   Table pour modele A1   %%%%%%%%%%%%%%%%
%\squeezetable  
\begin{table} \begin{center} 
\caption{ \label{dirplatesa1} Superpotential couplings for model $A1$
of bilinear order in $h_i , \ C_i , q_i, \ \bar q_i $ and orders $ n
\leq 4$ in the singlets.  The column entries refer to the composite
flat direction $\calp ^{(2)} $ and the four randomly selected
individual flat directions: $P^{(3)} _{I}= Y_ {12} Y_ {16} Y_ {19}^2
,\ P^{(4)}_{II} = S_ { 1} S_ { 3} Y_ {16} Y_ {19}^2,\ P^{(4)} _{
III}=S_ { 1} S_ { 3}^2S_ { 4} S_ { 6}^2,\ P_{IV}^{(4)} =S_ { 1} S_ {
4} Y_ {18}^2 .$
%(The indices refer  to  the orders in personal lists.) 
Empty entries correspond to cases where no coupling is present up to
order $ n=4$.}
\vskip 0.5 cm \begin{tabular}{|c|c||cccc|} \hline $ W $ & $ \calp
^{(2)} $ & $ P^{(3)} _{I} $ & $ P^{(4)} _{II} $ & $ P^{(4)}_{ III} $ &
$ P^{(4)}_{IV} $ \\ \hline $ h_1 h_1 $ & $ \phi ^ 5Y_ {17} +\phi ^ 5S_
{ 4} $ & & & & \\ $ h_2 h_2 $ & & & $ \phi ^5S_ {13} $ & & \\ $ h_3
h_3 $ & $S_ { 1}+\phi S_ {15} +\phi ^ 5S_ { 7} $ & $ \phi S_ {11} + S_
{ 1} $ & $\phi +\phi ^2Y_ {12} $ & $\phi $ & $\phi +\phi ^2Y_ {14} $
\\ $ h_1 h_2 $ & $ S_ {10}+\phi ^ 2S_ {10} $ & $ S_ {10}$ & $ S_ {10}
$ & $ S_ {10} $ & $ S_ {10} $ \\ $ h_1 h_3 $ & $ \phi ^ 4 + S_ { 9} +
\phi ^ 3S_ {11} $ & $ S_ { 9} $ & $ S_ { 9}+\phi ^5Y_ { 6} $ & $ S_ {
9} $ & $ S_ { 9}+\phi ^5Y_ {11} $ \\ $ h_2 h_3 $ & $\phi ^ 6 +\phi ^
4S_ { 9} $ & & & $\phi ^4Y_ {12} +\phi ^4Y_ {14} $ & \\ \hline $ C_1
C_1 $ & $ \phi ^ 3Y_ {10} $ & & $ \phi ^ 5Y_ {3} $ & & $ \phi ^5Y_ {
3} $\\ $ C_2 C_2 $ & $\phi ^ 4S_ {10} $ & & $ \phi ^5S_ {13} $ & & \\
$ C_3 C_3 $ & $ S_ { 1} + \phi S_ {15} + \phi ^ 5S_ { 7} $ & $\phi S_
{11} + S_ { 1} $ & $\phi +\phi ^2Y_ {12} $ & $ \phi $ & $ \phi +\phi
^3Y_ {14} $ \\ $ C_4 C_4 $ & $ \phi ^ 4Y_ { 7} $ & & $\phi ^ 3 Y_ { 5}
$ & & $\phi ^3Y_ { 9} $ \\ $ C_1 C_2 $ & & & & & \\ $ C_1 C_3 $ & $
\phi ^ 2Y_ { 3} +\phi ^ 4Y_ { 4} $ & & $\phi ^3Y_ { 3} $ & & $ \phi
^3Y_ { 3} $ \\ $ C_1 C_4 $ & & & & & \\ $ C_2 C_3 $ & $ \phi ^ 6 +\phi
^ 4S_ { 9} $ & & & $ \phi ^4Y_ {12} +\phi ^4Y_ {14} $ & \\ $ C_2 C_4 $
& $\phi ^ 5Y_ { 9} $ & & & & \\ $ C_3 C_4 $ & & & $ \phi^ 2 Y_ { 5}$ &
& $\phi ^2 Y_{ 9}$ \\ \hline $ q_1 \bar q_1 $ & & & & & \\ $ q_1 \bar
q_2 $ & $ \phi ^ 3S_ { 9} +\phi ^5 $ & & & $ \phi ^4Y_ {14} +\phi ^4Y_
{12} $ & \\ $ q_2 \bar q_1 $ & & & & & \\ $q_2 \bar q_2 $ & $ S_ {14}
$ & $ \phi^ 2 S_{14} $ & $S_ {14} $ & $S_ {14} $ & $S_ {14} $ \\
\hline
\end{tabular} \end{center}  \end{table} 

To complete the present study, we have scanned the $70$ and $48$ flat
monomial solutions of orders $n=3$ and $ n= 4 $ in search of the
bidoublets couplings of form, $ W_h= h_i h_j ( \mu _{ij} + \l _{ij}^k
\phi ^k ) $.  The global results combining the contributions from the
various individual flat directions read: \bea \bullet \ n=3:\ W_h &=&
h_1 h_1 [ 0 ] + h_2 h_2 [ \phi ^ 4 S_ {10}] + h_3 h_3 [ \phi S_ { 11}]
+ h_1 h_2 [\phi ^ 2S_{13} ] + h_1 h_3 [\phi ^ 2Y_{15}] + h_2 h_3 [
\phi ^6+ \phi ^ 4 Y_ {15}] . \cr \bullet \ n=4:\ W_h &=& h_1 h_1 [ 0 ]
+ h_2 h_2 [ \phi ^ 5 S_ {12,13}] + h_3 h_3 [ \phi +\phi ^ 2Y_ {
12,14}] + h_1 h_2 [ S_{10} + \phi ^ 2Y_ {6, 11} ] \cr && + h_1 h_3
[S_9 + \phi ^ 5Y_ { 6}] + h_2 h_3 [\phi ^ 4 Y_ {12,14}] .  \eea
Happily, it appears that the cases with no trilinear couplings, $ \mu
_{ij} \ne 0 , \ \l _{ij}^k = 0$, are outnumbered by those with no
bilinear couplings, $ \mu _{ij} = 0 , \ \l _{ij}^k \ne 0$.  The
results for the directions $ n=3$ favor a scenario with a single light
bidoublet $h_1$ having trilinear couplings, $\phi ^ 2 h_1 h_2 S_{13} $
or $\phi ^ 2 h_1 h_3 Y_{15} $.  The results for the directions $n=4$
favor three light bidoublets with canonical and non-canonical type
trilinear couplings.

%%%%%%%%%%%%%%%%%%%%%%%%%%%%%%%%%%%%%%%%%

%\setcounter{table}{0} 

\subsection{Model $A2$}

\squeezetable \begin{table}[H] \begin{center}
\caption{ \label{tabmdta2} String spectrum of massless modes for model
$ A2$ of~\cite{Kobayashi:2004ya} of gauge group, $ G_{422} \times
(SO(10)'\times SU(2) ') \times U(1)^5$, with the pair of group
factors, $SU(2)_{1,2} $, identified with $ SU(2)_{L,R} $. Same
notational conventions as in Table~\ref{tabmoda1} are used. The
Abelian charges (given by integers except for $ Q_4$) are defined in
the rotated charge basis by the rescalings, $ Q_1/0.1667, \ Q_2
/0.1667 \ , \ 4 \ Q_3 /0.00260 , \ Q_4/ 0.0007667 , \ 18 \
Q_A/0.1244.$}
\vskip 0.5 cm 
\begin{tabular}{|ccccccccccccccc||cccccccccccccc|}\hline
$g$ & M & $N_{32}$ & $\gamma$ & $\cald$ && $ 6 (R_1$ & $R_2$ & $R_3 ) $ && $Q_1$ & $Q_2$ & $Q_3$ & $Q_4$ & $Q_A$ & M & $N_{32}$ & $\gamma$ & $\cald$ && $ 6 (R_1$ & $R_2$ & $R_3) $ && $Q_1$ & $Q_2$ & $Q_3$ & $Q_4$ & $Q_A$ \\   \hline 
  0 & $\S '_{ 1}$ & $1$ & $1$ & $1$ && $6$ & $0$ & $0$ && $0$ & $0$ & $-18$ & $15.05$ & $-432$
  & $d'_{ 1}$ & $1$ & $1$ & $1$ && $6$ & $0$ & $0$ && $0$ & $0$ & $0$ & $-122.28$ & $-54$
 \\ 
  & $S_ { 1}$ & $1$ & $1$ & $1$ && $0$ & $6$ & $0$ && $0$ & $-6$ & $48$ & $0.63$ & $-18$
  & $f_ { 1}$ & $1$ & $1$ & $1$ && $6$ & $0$ & $0$ && $3$ & $3$ & $24$ & $0.31$ & $-9$
 \\ 
 \hline 
  1 & $Y_ { 1}$ & $1$ & $1$ & $2$ && $1$ & $2$ & $9$ && $2$ & $2$ & $-20$ & $-10.45$ & $-96$
  & $Y_ { 2}$ & $1$ & $1$ & $2$ && $1$ & $2$ & $-3$ && $2$ & $2$ & $-20$ & $-10.45$ & $-96$
 \\ 
  & $Y_ { 3}$ & $1$ & $1$ & $2$ && $-5$ & $-4$ & $3$ && $2$ & $2$ & $-20$ & $-10.45$ & $-96$
  & $Y_ { 4}$ & $1$ & $1$ & $2$ && $-17$ & $2$ & $3$ && $2$ & $2$ & $-20$ & $-10.45$ & $-96$
 \\ 
  & $Y_ { 5}$ & $1$ & $1$ & $2$ && $-5$ & $2$ & $3$ && $2$ & $-4$ & $28$ & $-9.82$ & $-114$
  & $Y_ { 6}$ & $1$ & $1$ & $2$ && $-5$ & $2$ & $3$ && $-4$ & $-4$ & $-20$ & $-10.45$ & $-96$
 \\ 
  & $Y_ { 7}$ & $1$ & $1$ & $2$ && $-5$ & $2$ & $3$ && $-4$ & $2$ & $28$ & $-9.82$ & $-114$
  & $f^c_{ 1}$ & $1$ & $1$ & $2$ && $1$ & $2$ & $3$ && $-1$ & $-1$ & $4$ & $-10.14$ & $-105$
 \\ 
  & $f_ { 2}$ & $1$ & $1$ & $2$ && $1$ & $2$ & $3$ && $-1$ & $-1$ & $4$ & $-10.14$ & $-105$
  & $Y_ { 8}$ & $2$ & $1$ & $2$ && $-5$ & $2$ & $3$ && $0$ & $0$ & $12$ & $-91.55$ & $-144$
 \\ 
  & $\S '_{ 2}$ & $2$ & $1$ & $2$ && $1$ & $2$ & $3$ && $0$ & $0$ & $18$ & $-15.05$ & $36$
  & $d'_{ 2}$ & $2$ & $1$ & $2$ && $1$ & $-4$ & $3$ && $0$ & $0$ & $12$ & $30.73$ & $-90$
 \\ 
  & $d'_{ 3}$ & $2$ & $1$ & $2$ && $-11$ & $2$ & $3$ && $0$ & $0$ & $12$ & $30.73$ & $-90$
  & $Y_ { 9}$ & $2$ & $1$ & $2$ && $-5$ & $2$ & $3$ & &$0$ & $0$ & $36$ & $-30.10$ & $468$
 \\ 
  & $d'_{ 4}$ & $3$ & $1$ & $2$ && $1$ & $2$ & $3$ && $4$ & $-2$ & $-4$ & $-51.00$ & $-120$
  & $Y_ {10}$ & $3$ & $1$ & $2$ && $-5$ & $2$ & $3$ && $4$ & $-2$ & $-4$ & $71.28$ & $-66$
 \\ 
  & $d'_{ 5}$ & $3$ & $1$ & $2$ && $1$ & $2$ & $3$ && $-2$ & $4$ & $-4$ & $-51.00$ & $-120$
  & $Y_ {11}$ & $3$ & $1$ & $2$ && $-5$ & $2$ & $3$ && $-2$ & $4$ & $-4$ & $71.28$ & $-66$
 \\ 
  & $\bar q_{ 1}$ & $4$ & $1$ & $2$ && $-5$ & $2$ & $3$ && $-1$ & $2$ & $-20$ & $-10.45$ & $-96$
  & $q_ { 1}$ & $4$ & $1$ & $2$ && $1$ & $2$ & $3$ && $-1$ & $-4$ & $-20$ & $-10.45$ & $-96$
 \\ 
  & $q_ { 2}$ & $4$ & $1$ & $2$ && $1$ & $2$ & $3$ && $-1$ & $2$ & $28$ & $-9.82$ & $-114$
  & $d^l_{ 1}$ & $4$ & $1$ & $2$ && $1$ & $2$ & $3$ && $2$ & $-1$ & $-44$ & $-10.76$ & $-87$
 \\ 
  & $d^l_{ 2}$ & $4$ & $1$ & $2$ && $1$ & $2$ & $3$ && $2$ & $5$ & $4$ & $-10.14$ & $-105$
  & $d^r_{ 1}$ & $4$ & $1$ & $2$ && $1$ & $-4$ & $3$ && $2$ & $-1$ & $4$ & $-10.14$ & $-105$
 \\ 
  & $d^r_{ 2}$ & $4$ & $1$ & $2$ && $-11$ & $2$ & $3$ && $2$ & $-1$ & $4$ & $-10.14$ & $-105$
  & $d^l_{ 3}$ & $4$ & $1$ & $2$ && $-5$ & $2$ & $3$ && $-4$ & $-1$ & $4$ & $-10.14$ & $-105$
 \\ 
  & $h ^{'r}_{ 1}$ & $5$ & $1$ & $2$ && $1$ & $2$ & $3$ && $0$ & $3$ & $-12$ & $30.41$ & $-81$
  & $h ^{'l} _{ 1}$ & $5$ & $1$ & $2$ && $1$ & $2$ & $3$ && $0$ & $-3$ & $-12$ & $30.41$ & $-81$
 \\ 
  & $\bar q_{ 2}$ & $6$ & $1$ & $2$ && $1$ & $2$ & $3$ && $1$ & $-2$ & $-4$ & $71.28$ & $-66$
  & $d^l _{ 4}$ & $6$ & $1$ & $2$ && $1$ & $2$ & $3$ && $-2$ & $1$ & $-28$ & $70.97$ & $-57$
 \\ 
  & $d^r_{ 3}$ & $6$ & $1$ & $2$ && $1$ & $2$ & $3$ && $-2$ & $1$ & $20$ & $71.59$ & $-75$
 &&&&&&&&&&&&&& \\  \hline 
  2 & $S_ { 2}$ & $1$ & $-1$ & $1$ && $2$ & $4$ & $0$ && $4$ & $4$ & $-40$ & $-20.90$ & $-192$
  & $Y_ {12}$ & $1$ & $1$ & $2$ && $-4$ & $4$ & $0$ && $4$ & $-2$ & $8$ & $-20.28$ & $-210$
 \\ 
  & $Y_ {13}$ & $1$ & $-1$ & $1$ && $2$ & $10$ & $0$ && $4$ & $-2$ & $8$ & $-20.28$ & $-210$
  & $S_ { 3}$ & $1$ & $-1$ & $1$ && $2$ & $4$ & $0$ && $4$ & $-2$ & $32$ & $41.18$ & $402$
 \\ 
  & $C_ { 1}$ & $1$ & $-1$ & $1$ && $2$ & $4$ & $0$ && $-2$ & $-2$ & $8$ & $-20.28$ & $-210$
  & $f^c_{ 2}$ & $1$ & $1$ & $2$ && $2$ & $4$ & $0$ && $1$ & $1$ & $-16$ & $-20.59$ & $-201$
 \\ 
  & $d'_{ 6}$ & $2$ & $-1$ & $2$ && $2$ & $4$ & $0$ && $-4$ & $-4$ & $-8$ & $20.28$ & $-186$
  & $S_ { 4}$ & $2$ & $1$ & $1$ && $2$ & $4$ & $0$ && $-4$ & $-4$ & $16$ & $-40.55$ & $372$
 \\ 
  & $S_ { 5}$ & $3$ & $-1$ & $1$ && $2$ & $4$ & $0$ && $0$ & $6$ & $-24$ & $60.83$ & $-162$
  & $h_ { 1}$ & $3$ & $-1$ & $1$ && $2$ & $4$ & $0$ && $0$ & $0$ & $-24$ & $60.83$ & $-162$
 \\ 
  & $S_ { 6}$ & $3$ & $-1$ & $1$ && $2$ & $4$ & $0$ && $0$ & $-6$ & $-24$ & $60.83$ & $-162$
  & $d'_{ 7}$ & $3$ & $1$ & $2$ && $2$ & $4$ & $0$ && $0$ & $0$ & $24$ & $-60.83$ & $-234$
 \\ 
  & $T'_{ 1}$ & $3$ & $-1$ & $1$ && $2$ & $4$ & $0$ && $0$ & $0$ & $36$ & $-30.10$ & $72$
  & $Y_ {14}$ & $3$ & $1$ & $2$ && $-4$ & $4$ & $0$ && $0$ & $0$ & $24$ & $61.45$ & $-180$
 \\ 
  & $Y_ {15}$ & $3$ & $-1$ & $1$ && $2$ & $10$ & $0$ && $0$ & $0$ & $24$ & $61.45$ & $-180$
  & $d'_{ 8}$ & $3$ & $1$ & $2$ && $2$ & $4$ & $0$ && $0$ & $0$ & $48$ & $0.63$ & $378$
 \\ 
 \hline 
  3 & $S_ { 7}$ & $1$ & $\o$ & $2$ && $3$ & $0$ & $3$ && $6$ & $0$ & $-12$ & $-30.73$ & $-306$
  & $S_ { 8}$ & $1$ & $\o ^2$ & $2$ && $3$ & $0$ & $3$ && $0$ & $0$ & $-36$ & $30.10$ & $324$
 \\ 
  & $S_ { 9}$ & $1$ & $\o$ & $2$ && $3$ & $0$ & $3$ && $0$ & $6$ & $-12$ & $-30.73$ & $-306$
  & $S_ {10}$ & $1$ & $\o$ & $2$ && $3$ & $0$ & $3$ && $0$ & $-6$ & $12$ & $30.73$ & $306$
 \\ 
  & $S_ {11}$ & $1$ & $1$ & $4$ && $3$ & $0$ & $3$ && $0$ & $0$ & $36$ & $-30.10$ & $-324$
  & $S_ {12}$ & $1$ & $\o$ & $2$ && $3$ & $0$ & $3$ && $-6$ & $0$ & $12$ & $30.73$ & $306$
 \\ 
  & $\bar q_{ 3}$ & $2$ & $\o ^2$ & $2$ && $3$ & $0$ & $3$ && $3$ & $0$ & $12$ & $30.73$ & $306$
  & $q_ { 3}$ & $2$ & $\o ^2$ & $2$ && $3$ & $0$ & $3$ && $-3$ & $0$ & $-12$ & $-30.73$ & $-306$
 \\ 
  & $d^r_{ 4}$ & $2$ & $\o$ & $4$ && $3$ & $0$ & $3$ && $0$ & $-3$ & $-12$ & $30.41$ & $315$
  & $d^r _{ 5}$ & $2$ & $1$ & $2$ && $3$ & $0$ & $3$ && $0$ & $3$ & $12$ & $-30.41$ & $-315$
 \\ 
 \hline 
  4 & $C_ { 2}$ & $1$ & $1$ & $2$ && $4$ & $2$ & $0$ && $2$ & $2$ & $-8$ & $20.28$ & $210$
  & $S_ {13}$ & $1$ & $1$ & $2$ && $4$ & $2$ & $0$ && $-4$ & $2$ & $-32$ & $-41.18$ & $-402$
 \\ 
  & $Y_ {16}$ & $1$ & $1$ & $2$ && $4$ & $-4$ & $0$ && $-4$ & $2$ & $-8$ & $20.28$ & $210$
  & $Y_ {17}$ & $1$ & $-1$ & $1$ && $10$ & $2$ & $0$ && $-4$ & $2$ & $-8$ & $20.28$ & $210$
 \\ 
  & $S_ {14}$ & $1$ & $1$ & $2$ && $4$ & $2$ & $0$ && $-4$ & $-4$ & $40$ & $20.90$ & $192$
  & $\bar f^c_{ 1}$ & $1$ & $-1$ & $1$ && $4$ & $2$ & $0$ && $-1$ & $-1$ & $16$ & $20.59$ & $201$
 \\ 
  & $d'_{ 9}$ & $2$ & $1$ & $1$ && $4$ & $2$ & $0$ && $0$ & $0$ & $-48$ & $-0.63$ & $-378$
  & $T'_{ 2}$ & $2$ & $-1$ & $2$ && $4$ & $2$ & $0$ && $0$ & $0$ & $-36$ & $30.10$ & $-72$
 \\ 
  & $Y_ {18}$ & $2$ & $-1$ & $2$ && $4$ & $-4$ & $0$ && $0$ & $0$ & $-24$ & $-61.45$ & $180$
  & $Y_ {19}$ & $2$ & $1$ & $1$ && $10$ & $2$ & $0$ && $0$ & $0$ & $-24$ & $-61.45$ & $180$
 \\ 
  & $d'_{10}$ & $2$ & $1$ & $1$ && $4$ & $2$ & $0$ && $0$ & $0$ & $-24$ & $60.83$ & $234$
  & $S_ {15}$ & $2$ & $-1$ & $2$ && $4$ & $2$ & $0$ && $0$ & $6$ & $24$ & $-60.83$ & $162$
 \\ 
  & $h_ { 2}$ & $2$ & $-1$ & $2$ && $4$ & $2$ & $0$ && $0$ & $0$ & $24$ & $-60.83$ & $162$
  & $S_ {16}$ & $2$ & $-1$ & $2$ && $4$ & $2$ & $0$ && $0$ & $-6$ & $24$ & $-60.83$ & $162$
 \\ 
  & $S_ {17}$ & $3$ & $1$ & $2$ && $4$ & $2$ & $0$ && $4$ & $4$ & $-16$ & $40.55$ & $-372$
  & $d'_{11}$ & $3$ & $-1$ & $1$ && $4$ & $2$ & $0$ && $4$ & $4$ & $8$ & $-20.28$ & $186$
 \\     \hline \hline
\end{tabular} \end{center}  \end{table}

 The mass spectrum of model $A2$  is displayed
in Table~\ref{tabmdta2}.  The D flatness conditions are solved at
order $n=2$ by the single monomial, $ Y_ { 9} S_ { 8} $, and at order
$n=3$ by the 7 monomials, $S_ { 1} S_ { 8}^2S_ {15} , \ Y_ { 9} S_ {
8}^2S_ {11} , \ S_ { 3} Y_ {16} Y_ {18} , \ S_ { 3} Y_ {17} Y_ {18} ,
\ S_ { 3} Y_ {17} Y_ {19} , \ S_ { 3}Y_ {16}Y_ {19}, \ S_ { 8} S_ {10}
S_{15} $.  The number of solutions at order $n=4$ exceeds $O(100)$.
The combined composite flat direction $ \calp ^{(2)} \times \calp
^{(3)} $ for $ n\leq 3 $ includes the eleven modes, $ Y_ { 9} , \ S_ {
8}, \ S_ { 1}, \ S_ {15}, \ S_{11} , \ S_ { 3}, \ Y_ {16}, \ Y_ {18},
\ Y_ {19}, \ Y_ {17}, \ S_ {10} .$ This remains unlifted by all the
singlet couplings in $ W_s$ of order $ n\leq 4 $.  However, no
unbroken $U(1) $ charges survive if the above eleven modes
simultaneously acquire finite VEVs.  The effective superpotential
component $W_s $ and those of bilinear and trilinear couplings
obtained by assigning a common VEV $\phi $ to the modes $\hat \phi _i
\in \calp ^{(2)} \times \calp ^{(3)} $ are given by \bea W_s &=& [\phi
^3 S_ { 9} ] + [\phi S_ { 1} S_ { 9} +\phi ^2 (Y_ { 1}^2 + Y_ { 2}^2 +
Y_ { 4}^2 ) +\phi ^2 Y_ { 6} S_ {17} +\phi ^2 S_ {13} S_ {17} +\phi ^2
Y_ {10} S_ {13} ] .  \cr && \cr W_h &=& h_1 h_1 [ \phi ^ 3+\phi ^ 3Y_
{ 5} ] + h_2 h_2 [ 0 ] + h_1 h_2 [ \phi ^ 2Y_ { 3} ] . \cr && \cr W_C
&=& C_1 C_1 [ \phi ^ 2Y_ { 3} ] + C_2 C_2 [\phi Y_ { 6} ] + C_1 C_2 [
\phi ^ 2Y_ { 3} ] . \cr && \cr W_q &=& q_1 \bar q_1 [\phi ^ 3Y_ { 1}
+\phi ^ 3Y_ { 2} ] + q_2 \bar q_2 [\phi ^ 3Y_ { 8} ] + q_3 \bar q_3 [
\phi Y_ { 1}Y_ { 7} +\phi Y_ { 2}Y_ { 7}] \cr && + q_1 \bar q_2 [\phi
+\phi^ 2S_ { 9} +\phi^ 2Y_ {12}] + q_1 \bar q_3 [\phi^ 4Y_ { 4}] + q_2
\bar q_3 [0] +\bar q_1 q_2 [0] + \bar q_1 q_3 [\phi^ 3S_ { 9}S_ {17} ]
+ \bar q_2 q_3 [\phi^ 2Y_ { 1}Y_ { 7} ] . \cr && \cr W_f &=& f^c_1
\bar f^c_1 [\phi Y_ { 3} (Y_ {12}+ Y_ {13} + S_ {13}) ] + f^c_2 \bar
f^c_1 [ 0 ] .  \cr && \cr W_{hff^c} &=& h_1 f_1 f^c_1 [ Y_ { 3}S_ { 4}
] + h_1 f_2 f^c_2 [ 0 ] + h_1 f_1 f^c_2 [ 0 ] + h_1 f_2 f^c_1 [0] +
h_2 f_1 f^c_1 [0 ] + h_2 f_2 f^c_2 [\phi Y_ {10} ] + h_2 f_1 f^c_2 [0
] \cr &+& h_2 f_2 f^c_1 [ 0 ] . \cr && \cr W_{Cff} &=& C_1 f^c_1f^c_1 [ 0] +
C_1 f^c_2f^c_2 [ \phi ^ 3Y_ {14} ] + C_2 f^c_1 f^c_1 [\phi Y_ {12} ] +
C_2 f^c_2 f^c_2 [\phi^ 2Y_ { 5}].
\label{eqa23}
\eea We see that the singlet superpotential $W_s$ has no constant term
but includes a harmless tadpole for the massive field, $ S_{ 9} $.
The massive singlets are $ S_ { 1,9,13, 17 } $ and $ Y_ { 1,2,4,6,10}
$.  The Higgs bidoublet couplings in $ W_h$ show that $ h_1 $
decouples, while $ h_2 $ is left massless but lacks trilinear
couplings to singlets.  Indeed, the quartic coupling, $ \phi Y_ {10}S_
{13} \ h_2 h_2 $, cannot generate an effective trilinear coupling at
lower scales, since both of the singlets $ Y_ {10},\ S_ {13}$ are
massive.  No large mass pairing takes place between the conjugate
bifundamental modes. The suppressed couplings in $W_{hff^c} $ preclude
the possibility of discriminating between the light and heavy flavors.
Some zero entries may be possibly lifted by assuming a finite VEV for
the singlet $ Y_{10}$.  The sextet and quartet modes, except the pair,
$ q_1 \bar q_2 $, remain massless.  The couplings in $ W_{Cff} $ give
no contributions to baryon number violating operators.

We have also scanned the individual D flat directions of orders $
n=2,\ 3 $ and $n=4$ in search of the bidoublet modes couplings.  The
full set of 15 monomials at orders $ n=2, \ 3 $ are F flat. Of the
$100$ monomials at order $n=4$, about 80 are lifted by F-terms.  The
results obtained by combining the contributions from the various flat
directions are \bea && \bullet \ n=2,\ 3:\ W_h = h_1 h_1 [ Y_ {18} ] +
h_2 h_2 [ 0] + h_1 h_2 [0 ] .  \cr && \bullet \ n=4:\ W_h = h_1 h_1 [
\phi ^4 Y_ {18,19} ] + h_2 h_2 [ \phi ^4 Y_ {14, 15} +\phi ^3 S_ {13}]
+ h_1 h_2 [\phi ^3 S_ { 1} +\phi ^2 Y_ { 5} +\phi ^4 S_ { 2} ] . \eea
We see again a clear trend in favor of several light bidoublets
coupled by trilinear terms to massless singlets.

%%%%%%%%%%%%%%%%%%%%%%%%%%%%%%%%%%%%%%%%

%\setcounter{table}{0} 

\subsection{Model $B$}
\squeezetable \begin{table}[H] \begin{center}
\caption{ \label{tabmodeleb} Massless string modes for model $B$
  of~\cite{Kobayashi:2004ya} with gauge group, $ G_{422} \times
  (SO(10)') \times U(1)^6$.  We use same notational conventions as in
  Table~\ref{tabmoda1}, except that the pair of group factor
  $SU(2)_{1,2} $ identify here with $ SU(2)_{R,L} $. The quoted
  integer rescaled charges are related to the initial ones by the
  rescalings, $ 4 \ Q_1/ 0.005208 , \ 25 \ Q_2/0.002042 \ , \ 12 \
  Q_3/ 0.009615 , \ 9 \ Q_4/0.00347, \ Q_5 / 0.000672 , \ Q_A/0.017796
  .$}
\vskip 0.5 cm 
\begin{tabular}{|cccccccccccccccc||ccccccccccccccc|}\hline 
$g$ & M & $N_{32}$ & $\gamma$ & $\cald$ && $ 6 (R_1$ & $R_2$ & $R_3 )$ && $Q_1$ & $Q_2$ & $Q_3$ & $Q_4$ & $Q_5$ & $Q_A$ & M & $N_{32}$ & $\gamma$ & $\cald$ && $ 6 (R_1$ & $R_2$ & $R_3)$ && $Q_1$ & $Q_2$ & $Q_3$ & $Q_4$ & $Q_5$ & $Q_A $ \\   \hline 
  0 & $S_ { 1}$ & $1$ & $1$ & $1$ && $0$ & $6$ & $0$ && $48$ & $210$ & $-12$ & $-24$ & $-12$ & $-12$
  & $T'_{ 1}$ & $1$ & $1$ & $1$ && $0$ & $6$ & $0$ && $0$ & $0$ & $0$ & $-81$ & $6$ & $6$
 \\ 
  & $C_ { 1}$ & $1$ & $1$ & $1$ && $0$ & $6$ & $0$ && $-12$ & $-180$ & $-12$ & $-24$ & $-12$ & $-12$
  & $f^c_{ 1}$ & $1$ & $1$ & $1$ && $0$ & $0$ & $6$ && $30$ & $195$ & $0$ & $0$ & $0$ & $0$
 \\ 
  & $f^c_{ 2}$ & $1$ & $1$ & $1$ && $6$ & $0$ & $0$ && $-18$ & $-15$ & $12$ & $24$ & $12$ & $12$
  & $\bar f^c _ { 1}$ & $1$ & $1$ & $1$ && $0$ & $0$ & $6$ && $-30$ & $-195$ & $0$ & $0$ & $0$ & $0$
 \\   \hline 
  1 & $Y_ { 1}$ & $1$ & $1$ & $2$ && $1$ & $-4$ & $3$ && $16$ & $155$ & $32$ & $-17$ & $7$ & $7$
  & $Y_ { 2}$ & $1$ & $1$ & $2$ && $-11$ & $2$ & $3$ && $16$ & $155$ & $32$ & $-17$ & $7$ & $7$
 \\ 
  & $S_ { 2}$ & $1$ & $1$ & $2$ && $1$ & $2$ & $3$ && $16$ & $155$ & $-46$ & $70$ & $4$ & $4$
  & $h_ { 1}$ & $1$ & $1$ & $2$ && $1$ & $2$ & $3$ && $4$ & $-25$ & $20$ & $-41$ & $-5$ & $-5$
 \\ 
   & $C_ { 2}$ & $1$ & $1$ & $2$ && $1$ & $2$ & $3$ && $4$ & $-25$ & $20$ & $-41$ & $-5$ & $-5$
   & $Y_ { 3}$ & $1$ & $1$ & $2$ && $-5$ & $2$ & $3$ && $4$ & $-280$ & $29$ & $-23$ & $4$ & $4$
 \\ 
   & $Y_ { 4}$ & $1$ & $1$ & $2$ && $-5$ & $2$ & $3$ && $-44$ & $20$ & $23$ & $-35$ & $-2$ & $-2$
   & $S_ { 3}$ & $2$ & $1$ & $2$ && $1$ & $2$ & $3$ && $32$ & $55$ & $34$ & $-13$ & $71$ & $7$
 \\ 
   & $S_ { 4}$ & $2$ & $1$ & $2$ && $1$ & $2$ & $3$ && $32$ & $55$ & $34$ & $68$ & $-28$ & $4$
   & $S_ { 5}$ & $2$ & $1$ & $2$ && $1$ & $2$ & $3$ && $32$ & $55$ & $-44$ & $-7$ & $-19$ & $13$
 \\ 
  & $Y_ { 5}$ & $2$ & $1$ & $2$ && $-5$ & $2$ & $3$ && $-28$ & $-80$ & $25$ & $-31$ & $62$ & $-2$
  & $Y_ { 6}$ & $2$ & $1$ & $2$ && $-5$ & $2$ & $3$ && $-28$ & $-80$ & $25$ & $50$ & $-37$ & $-5$
 \\ 
  & $Y_ { 7}$ & $2$ & $1$ & $2$ && $-5$ & $2$ & $3$ && $-28$ & $-80$ & $-53$ & $-25$ & $-28$ & $4$
  & $S_ { 6}$ & $3$ & $1$ & $2$ && $1$ & $2$ & $3$ && $0$ & $255$ & $30$ & $-21$ & $-57$ & $7$
 \\ 
  & $S_ { 7}$ & $3$ & $1$ & $2$ && $1$ & $2$ & $3$ && $0$ & $255$ & $30$ & $60$ & $30$ & $-2$
  & $S_ { 8}$ & $3$ & $1$ & $2$ && $1$ & $2$ & $3$ && $0$ & $255$ & $-48$ & $-15$ & $39$ & $7$
 \\ 
  & $Y_ { 8}$ & $3$ & $1$ & $2$ && $-5$ & $2$ & $3$ && $-12$ & $-180$ & $27$ & $-27$ & $-60$ & $4$
  & $Y_ { 9}$ & $3$ & $1$ & $2$ && $-5$ & $2$ & $3$ && $-12$ & $-180$ & $27$ & $54$ & $27$ & $-5$
 \\ 
  & $Y_ {10}$ & $3$ & $1$ & $2$ && $-5$ & $2$ & $3$ && $-12$ & $-180$ & $-51$ & $-21$ & $36$ & $4$
  & $d^l_{ 1}$ & $4$ & $1$ & $2$ && $1$ & $2$ & $3$ && $16$ & $155$ & $-7$ & $-14$ & $-38$ & $10$
 \\ 
  & $d^r_{ 1}$ & $4$ & $1$ & $2$ && $-5$ & $2$ & $3$ && $4$ & $-25$ & $-19$ & $-38$ & $-50$ & $-2$
  & $d^r_{ 2}$ & $4$ & $1$ & $2$ && $1$ & $2$ & $3$ && $4$ & $-25$ & $-19$ & $43$ & $37$ & $-11$
 \\ 
  & $\bar q_{ 1}$ & $4$ & $1$ & $2$ && $1$ & $2$ & $3$ && $-14$ & $-40$ & $-7$ & $-14$ & $-38$ & $10$
  & $d^l_{ 2}$ & $5$ & $1$ & $2$ && $1$ & $2$ & $3$ && $32$ & $55$ & $-5$ & $-10$ & $26$ & $10$
 \\ 
  & $d^r_{ 3}$ & $5$ & $1$ & $2$ && $1$ & $2$ & $3$ && $-16$ & $100$ & $-2$ & $-4$ & $29$ & $13$
  & $d^r_{ 4}$ & $5$ & $1$ & $2$ && $-5$ & $2$ & $3$ && $20$ & $-125$ & $-17$ & $-34$ & $14$ & $-2$
 \\ 
  & $d^l_{ 3}$ & $5$ & $1$ & $2$ && $-5$ & $2$ & $3$ && $-28$ & $-80$ & $-14$ & $-28$ & $17$ & $1$
  & $q_ { 1}$ & $5$ & $1$ & $2$ && $-5$ & $2$ & $3$ && $2$ & $115$ & $-14$ & $-28$ & $17$ & $1$
 \\ 
  & $\bar q_{ 2}$ & $5$ & $1$ & $2$ && $1$ & $2$ & $3$ && $2$ & $-140$ & $-5$ & $-10$ & $26$ & $10$
  & $d^r_{ 5}$ & $6$ & $1$ & $2$ && $1$ & $2$ & $3$ && $-12$ & $75$ & $-21$ & $39$ & $-27$ & $-11$
 \\ 
  & $d^l _{ 4}$ & $6$ & $1$ & $2$ && $1$ & $2$ & $3$ && $-12$ & $-180$ & $-12$ & $57$ & $-18$ & $-2$
  & $q_ { 2}$ & $6$ & $1$ & $2$ && $1$ & $2$ & $3$ && $18$ & $15$ & $-12$ & $57$ & $-18$ & $-2$
 \\  \hline 
  2 & $h_ { 2}$ & $1$ & $-1$ & $1$ && $2$ & $4$ & $0$ && $20$ & $130$ & $-26$ & $29$ & $-1$ & $-1$
  & $S_ { 9}$ & $1$ & $1$ & $2$ && $2$ & $4$ & $0$ && $20$ & $-125$ & $61$ & $-40$ & $11$ & $11$
 \\ 
  & $S_ {10}$ & $1$ & $1$ & $2$ && $2$ & $4$ & $0$ && $-28$ & $175$ & $55$ & $-52$ & $5$ & $5$
  & $S_ {11}$ & $1$ & $-1$ & $1$ && $2$ & $4$ & $0$ && $8$ & $-50$ & $40$ & $-82$ & $-10$ & $-10$
 \\ 
  & $f_{ 1}$ & $1$ & $1$ & $2$ && $2$ & $4$ & $0$ && $-10$ & $-65$ & $-26$ & $29$ & $-1$ & $-1$
  & $S_ {12}$ & $2$ & $1$ & $2$ && $2$ & $4$ & $0$ && $-12$ & $75$ & $57$ & $-48$ & $69$ & $5$
 \\ 
  & $Y_ {11}$ & $2$ & $1$ & $2$ && $2$ & $10$ & $0$ && $-12$ & $75$ & $-21$ & $-42$ & $-21$ & $11$
  & $Y_ {12}$ & $2$ & $-1$ & $1$ && $-4$ & $4$ & $0$ && $-12$ & $75$ & $-21$ & $-42$ & $-21$ & $11$
 \\ 
  & $T'_{ 2}$ & $2$ & $1$ & $2$ && $2$ & $4$ & $0$ && $-12$ & $75$ & $-21$ & $39$ & $-27$ & $5$
  & $S_ {13}$ & $2$ & $-1$ & $1$ && $2$ & $4$ & $0$ && $24$ & $-150$ & $-36$ & $-72$ & $-36$ & $-4$
 \\ 
  & $S_ {14}$ & $3$ & $1$ & $2$ && $2$ & $4$ & $0$ && $4$ & $-25$ & $59$ & $-44$ & $-53$ & $11$
  & $Y_ {13}$ & $3$ & $1$ & $2$ && $2$ & $10$ & $0$ && $4$ & $-25$ & $-19$ & $-38$ & $43$ & $11$
 \\ 
  & $Y_ {14}$ & $3$ & $-1$ & $1$ && $-4$ & $4$ & $0$ && $4$ & $-25$ & $-19$ & $-38$ & $43$ & $11$
  & $T'_{ 3}$ & $3$ & $1$ & $2$ && $2$ & $4$ & $0$ && $4$ & $-25$ & $-19$ & $43$ & $37$ & $5$
 \\ 
  &$\bar  \S '_{ 1}$ & $3$ & $1$ & $2$ && $2$ & $4$ & $0$ && $4$ & $-25$ & $20$ & $-1$ & $-8$ & $8$
  & $S_ {15}$ & $3$ & $-1$ & $1$ && $2$ & $4$ & $0$ && $-8$ & $50$ & $-40$ & $-80$ & $22$ & $-10$
 \\   \hline 
  3 & $Y_ {15}$ & $1$ & $1$ & $4$ && $-3$ & $0$ & $3$ && $24$ & $105$ & $-6$ & $-12$ & $-6$ & $-6$
  & $Y_ {16}$ & $1$ & $\o$ & $2$ && $9$ & $0$ & $3$ && $24$ & $105$ & $-6$ & $-12$ & $-6$ & $-6$
 \\ 
  & $Y_ {17}$ & $1$ & $\o ^2$ & $2$ && $3$ & $0$ & $9$ && $24$ & $105$ & $-6$ & $-12$ & $-6$ & $-6$
  & $Y_ {18}$ & $1$ & $\o ^2$ & $2$ && $3$ & $0$ & $-3$ && $24$ & $105$ & $-6$ & $-12$ & $-6$ & $-6$
 \\ 
  & $Y_ {19}$ & $1$ & $1$ & $4$ && $3$ & $0$ & $9$ && $-24$ & $-105$ & $6$ & $12$ & $6$ & $6$
  & $Y_ {20}$ & $1$ & $1$ & $4$ && $3$ & $0$ & $-3$ && $-24$ & $-105$ & $6$ & $12$ & $6$ & $6$
 \\ 
  & $Y_ {21}$ & $1$ & $\o$ & $2$ && $-3$ & $0$ & $3$ && $-24$ & $-105$ & $6$ & $12$ & $6$ & $6$
  & $Y_ {22}$ & $1$ & $\o ^2$ & $2$ && $9$ & $0$ & $3$ && $-24$ & $-105$ & $6$ & $12$ & $6$ & $6$
 \\ 
  & $f _{ 2}$ & $1$ & $1$ & $4$ && $3$ & $0$ & $3$ && $6$ & $90$ & $6$ & $12$ & $6$ & $6$
  & $f^c_{ 3}$ & $1$ & $1$ & $4$ && $3$ & $0$ & $3$ && $6$ & $90$ & $6$ & $12$ & $6$ & $6$
 \\ 
  & $\bar f _{ 1}$ & $1$ & $\o ^2$ & $2$ && $3$ & $0$ & $3$ && $-6$ & $-90$ & $-6$ & $-12$ & $-6$ & $-6$
  & $\bar f^c _ { 2}$ & $1$ & $\o ^2$ & $2$ && $3$ & $0$ & $3$ && $-6$ & $-90$ & $-6$ & $-12$ & $-6$ & $-6$
 \\ 
  & $d^r_{ 6}$ & $2$ & $\o$ & $4$ && $3$ & $0$ & $3$ && $24$ & $-150$ & $-36$ & $9$ & $-42$ & $6$
  & $d^r _{ 7}$ & $2$ & $1$ & $2$ && $3$ & $0$ & $3$ && $-24$ & $150$ & $36$ & $-9$ & $42$ & $-6$
 \\   \hline 
  4& $S_ {16}$ & $1$ & $1$ & $2$ && $4$ & $2$ & $0$ && $-8$ & $50$ & $-40$ & $82$ & $10$ & $10$
  & $S_ {17}$ & $1$ & $-1$ & $1$ && $4$ & $2$ & $0$ && $28$ & $-175$ & $-55$ & $52$ & $-5$ & $-5$
 \\ 
  & $S_ {18}$ & $1$ & $-1$ & $1$ && $4$ & $2$ & $0$ && $-20$ & $125$ & $-61$ & $40$ & $-11$ & $-11$
  & $h_ { 3}$ & $1$ & $1$ & $2$ && $4$ & $2$ & $0$ && $-20$ & $-130$ & $26$ & $-29$ & $1$ & $1$
 \\ 
  & $\bar f _{ 2}$ & $1$ & $-1$ & $1$ && $4$ & $2$ & $0$ && $10$ & $65$ & $26$ & $-29$ & $1$ & $1$
  & $S_ {19}$ & $2$ & $1$ & $2$ && $4$ & $2$ & $0$ && $8$ & $-50$ & $40$ & $80$ & $-22$ & $10$
 \\ 
  & $T'_{ 4}$ & $2$ & $-1$ & $1$ && $4$ & $2$ & $0$ && $-4$ & $25$ & $19$ & $-43$ & $-37$ & $-5$
  & $Y_ {23}$ & $2$ & $1$ & $2$ && $10$ & $2$ & $0$ && $-4$ & $25$ & $19$ & $38$ & $-43$ & $-11$
 \\ 
  & $Y_ {24}$ & $2$ & $-1$ & $1$ && $4$ & $-4$ & $0$ && $-4$ & $25$ & $19$ & $38$ & $-43$ & $-11$
  & $S_ {20}$ & $2$ & $-1$ & $1$ && $4$ & $2$ & $0$ && $-4$ & $25$ & $-59$ & $44$ & $53$ & $-11$
 \\ 
  & $\S '_{ 1}$ & $2$ & $-1$ & $1$ && $4$ & $2$ & $0$ && $-4$ & $25$ & $-20$ & $1$ & $8$ & $-8$
  & $S_ {21}$ & $3$ & $1$ & $2$ && $4$ & $2$ & $0$ && $-24$ & $150$ & $36$ & $72$ & $36$ & $4$
 \\ 
  & $T'_{ 5}$ & $3$ & $-1$ & $1$ && $4$ & $2$ & $0$ && $12$ & $-75$ & $21$ & $-39$ & $27$ & $-5$
  & $Y_ {25}$ & $3$ & $1$ & $2$ && $10$ & $2$ & $0$ && $12$ & $-75$ & $21$ & $42$ & $21$ & $-11$
 \\ 
   & $Y_ {26}$ & $3$ & $-1$ & $1$ && $4$ & $-4$ & $0$ && $12$ & $-75$ & $21$ & $42$ & $21$ & $-11$
 & $S_ {22}$ & $3$ & $-1$ & $1$ && $4$ & $2$ & $0$ && $12$ & $-75$ & $-57$ & $48$ & $-69$ & $-5$
 \\     \hline \hline 
\end{tabular}  \end{center}  \end{table}

The massless spectrum for model $B$   is  displayed  in
Table~\ref{tabmodeleb}. We find no holomorphic invariant monomials at order
$n=2$. At order $n=3$, there appears 16 solutions of which a few
representative monomials are: $ P_\a = [S_ {11} S_ {18} Y_ {25} , \ S_
{11} S_ {18} Y_ {26} , \ S_ {11} Y_ {23} S_ {20} , \ S_ {11} Y_ {24}
S_ {20} ]. $ The composite direction, $\calp ^{(3)} $, consists of the
eight modes, $ S_ {11},\ S_ {18}, \ Y_ {25}, \ Y_ {26}, \ Y_{23}, \ S_
{20}, \ Y_ {24}, \ S_ {15} $.  No Abelian charges remain unbroken
along $\calp ^{(3)} $.  Assigning a common VEV $\phi $ to these modes
yields the reduced superpotential 
\bea W_s &=& \phi Y_ { 3} S_ { 8}
+\phi Y_ { 4} S_ { 5} +\phi S_ { 3} Y_ { 7} +\phi S_ { 5} Y_ { 5}
+\phi S_ { 6} Y_ {10} +\phi S_ { 8} Y_ { 8}.\cr && \cr W_h &=& h_1
h_1 [ S_ {16} ] + h_2 h_2 [0]+ h_3 h_3 [0] + h_1 h_2 [ 0 ] + h_1 h_3 [
S_ { 2} ] + h_2 h_3 [ 0 ] .  \cr && \cr W_C &=& C_1 C_1 [0 ]+ C_2 C_2
[ S_ {16} ] + C_1 C_2 [ 0 ] . \cr && \cr W_q &=& q_1 \bar q_1 [ \phi ]
+ q_2 \bar q_2 [ 0 ] + q_1 \bar q_2 [ \phi ]+ q_2 \bar q_1 [0].  \cr
&& \cr W_f &=& f_1 \bar f_1 [ 0 ] + f_1 \bar f_2 [ 0 ] + f_2 \bar f_1
[0 ] + f_2 \bar f_2 [\phi Y_ { 3}+\phi Y_ { 8}] + f^c_1 \bar f^c_1 [
S_ { 1}(Y_ {19} + Y_ {20}) ] + f^c_1 \bar f^c_2 [0] 
\cr && + f^c_2\bar f^c_1 [ S_ { 1}] +
f^c_2 \bar f^c_2 [ S_ { 1}Y_ {21}] + f^c_3 \bar
f^c_1[ S_ { 1}Y_ {19}] + f^c_3 \bar f^c_2[\phi ^ 2Y_
{ 1}Y_ { 7}  ] . \cr && \cr W_{hff^c} &=& h_1
f_1 f^c_1 [Y_ {19} ] + h_1 f_1 f^c_2 [ Y_ {15} + Y_ {15}Y_ {19} ] +
h_1 f_1 f^c_3 [\phi Y_ {12} +\phi Y_ {14} ] \cr && + h_1 f_2 f^c_1 [ 0
] +h_1 f_2 f^c_2 [0] + h_1 f_2 f^c_3 [0 ] + h_2 f_1 f^c_1 [ 0 ] + h_2
f_1 f^c_2 [ 0 ] + h_2 f_1 f^c_3 [0 ] \cr && + h_2 f_2 f^c_1 [ 0 ]+ h_2
f_2 f^c_2 [ 0 ] + h_2 f_2 f^c_3 [ 0 ] + h_3 f_1 f^c_1 [\phi Y_ {12}
+\phi Y_ {14}] + h_3 f_1 f^c_2 [Y_ {15} ] + h_3 f_1 f^c_3 [Y_ {15} ]
\cr && + h_3 f_2 f^c_1 [ 0 ] + h_3 f_2 f^c_2 [0 ] + h_3 f_2 f^c_3 [0].  
\cr 
&& \cr 
W_{Cff} &=& C_1 f_1 f_1 [0] + C_1 f_2 f_2 [0] + C_1 f^c_1f^c_1 [   Y_  {19}  +   Y_  {20}]  
+ C_1 f^c_2f^c_2[Y_  {15} ] +  C_1   f^c_3 f^c_3  [\phi  Y_  {12}  +\phi  Y_  {14}] \cr 
&& + C_2 f_1 f_1 [0] + C_2 f_2f_2 [0]  + C_2 f^c_1f^c_1[0] + C_2 f^c_2f^c_2 [0] + 
C_2   f^c_3 f^c_3   [0 ] .  
\label{eqb1x1}  \eea
The identical couplings for $h_1$ and $C_2$ is a consequence of the
fact that these modes have same quantum numbers.  The singlets
superpotential $W_s$ contains no constant or linear terms, while
bilinear terms appear for the singlets $ S_ {3, 5,6, 8} $ and $
Y_{3,4,5,7,8,10}$, which thus pick up large masses.  The bidoublets
couplings in $ W_h$ show that the three sets of $h_i$ modes remain
massless, and have trilinear couplings $(hh S_{2} + hh S_{10}) $
involving the massless singlets $ S_2 $ and $ S_{10}$.  No mass
pairings arise between the conjugate bifundamental modes that would
remove some of the $f_i , \ f_i ^c $ modes or decouple some of the
mirror fermions, $ f_i , \ \bar f_i $ and $ f^c_i , \ \bar f^c_i $.
The bidoublet-fermions couplings in $ W_{hff^c} $ give vanishing mass
matrices for the matter fermions.  The flavor structure is problematic
since it gives no hint on how to discriminate between the heavy and
light flavors or between the matter and Higgs modes, independently of
the dominant linear combinations of the bidoublets.  The results for $
W_C$ and $ W_q$ show that all the sextet modes, $C_i$, and the single
pair of quartet modes, $ q_2 , \ \bar q_1$, remain unpaired.

The special role of the singlet $S_{16}$ motivates us in studying its
self-couplings at orders $ n \leq 4 $.  The results for $\calp ^{(3)}
: \ W_s (S_{16}) = S_{16} [0] + S_{16} ^2 [0] + S_{16} ^3 [0 ] , $
indicate that the mode $ S_{16} $ is light but has a suppressed cubic
coupling. The results for $\calp ^{(4)}, \ W_s ( S_{16} ) = S_{16} [0]
+ S_{16} ^2 [\phi^ 5Y_ {13} +\phi^ 5Y_ {11}] + S_{16} ^3 [\phi^ 4Y_ {
3} +\phi^ 4Y_ { 8}] $ would also allow for a light mode $ S_{16} $
provided that the massive singlets $Y_{3, 8, 11} $ acquire small VEVs.

Results for seven randomly selected individual flat directions are
displayed in Table~\ref{dirplatb}.  For all monomials other than $
P_{IV}$, we see that all the $ h_i$ remain light and have trilinear
couplings to singlets, but that the sextets and a subset of the
quartets fail to decouple.  We have also performed global type scans
of the D flat monomial solutions restricted to the bidoublets
couplings.  Of the 16 and $100 $ solutions at orders $ n=3 $ and $ 4
$, we find 0 and 10 monomials lifted by F-terms.  The bidoublets
couplings combining the contributions from the various flat directions
are given by \bea \bullet \ n=3:\ W_h &=& h_1 h_1 [ S_ {16} ] + h_2
h_2 [ 0] + h_3 h_3 [ 0] + h_1 h_2 [0 ] + h_1 h_3 [S_ { 2}] + h_2 h_3 [
0] .  \cr \bullet \ n=4:\ W_h &=& h_1 h_1 [ S_ {16} + \phi Y_ {19}
+\phi Y_ {20}] + h_2 h_2 [ 0] + h_3 h_3 [\phi Y_ {15} +\phi S_ { 2}] +
h_1 h_2 [ 0] \cr && + h_1 h_3 [ \phi + \phi S_ { 2} +\phi S_ {16}] +
h_2 h_3 [\phi ^3 Y_ {13}+ \phi ^2 S_ { 5} +\phi ^2 Y_ { 1}] .  \eea
These results again favor the scenario in which light bidoublets
interact by trilinear couplings to singlets.  A scan of the flat
directions reveals the existence of a large majority of individual
monomial directions along which the mode $S_{16}$ is light but with a
suppressed cubic self-coupling.

%\squeezetable  
\begin{table} \begin{center} 
\caption{   \label{dirplatb} 
Bilinear superpotential couplings  for model $B$ in  the
  modes  $h_i , \ C_i ,  q_i, \   \bar q_i $  of order, $ n  \leq 4$
  in the singlets.   The column entries  refer to  the   three flat directions of order    $ n =3:
\ P_{I}= S_  {11}  S_  {18}  Y_  {25}   , \   P_{II}=  S_  {11}  Y_
   {23}  S_  {20}   , \   P_{ III}=  S_  {15}  Y_  {23}  Y_  {26}   ,$
   and  the   four   randomly selected flat directions  of order    $ n =4:
\ P_{IV}=   S_  { 2}  Y_  { 6}  S_  {11}^2S_  {20}  , \  
P_{V}=  Y_  { 4}^2Y_  {15}^2S_  {17}  Y_  {26}   , \   P_{VI}=   Y_
  {6}  S_  {15}  Y_  {15}  Y_  {26}   , \  P_{VII}=  Y_  { 9}  S_
  {15}  Y_  {18}  Y_  {23}     $.  
Empty entries  correspond  to  cases where  no  coupling is present   
up to order $ n=4$.} 
\vskip  0.5 cm \begin{tabular}{|c|ccc||cccc|} \hline   
$   W  $  & $    P_I $  & $    P_{II} $
  & $   P_{III}  $  & $ P_{IV}  $  & $    P_{V}  $  &  $  P_{VI}  $ &
  $P_{VII}  $   \\  \hline  
 $  h_1  h_1 $  &    $S_  {16}$ &  $S_  {16}$ & $S_  {16}$ &  
$S_  {16}+\phi Y_  {19,20}$  & $S_  {16}$ &  $S_  {16}$ &  $S_  {16}$ \\  
 $   h_2 h_2 $ &        &       &         &        &      &       &       \\ 
  $ h_3 h_3 $ &           &       &       &     $\phi  Y_  {15}$ &  $\phi  S_  { 2}
 +\phi^2 S_  {16} $ &  $\phi   S_  { 2}  +\phi^2 S_  {16}$ &       \\  
  $   h_1  h_2 $   &      &        &       &      &       &       &      \\  
  $   h_1  h_3  $   & $S_  { 2}$  & $S_  { 2}$  &  $S_  { 2}$  & $\phi  $  &  $S_  { 2}
 +\phi   S_  {16}$ &  $S_  { 2}
 +\phi   S_  {16}$ & $S_  { 2}$ \\ 
 $  h_2  h_3 $  &          &        &        & $\phi^3 Y_  {13}$  & $\phi^2 Y_  { 1}
 +\phi^2  S_  { 5}$   &        &       \\ 
\hline 
$ C_1  C_1  $ &          &       &       &   &      &       &       \\ 
  $ C_2  C_2   $         &   $S_  {16}$ &  $S_  {16}$ & $S_  {16}$ & $S_  {16}
 +\phi   Y_  {19,20} $   &$S_  {16}$   & $S_  {16}$  &$S_  {16}$  \\ 
    $  C_1  C_2  $         &      &       &       &  & & & \\ 
 \hline  $   q_1   \bar q_1   $       &   $\phi $ & $Y_  {25}$  &   $Y_  {25}
 +\phi^2  Y_  {11,14}+\phi^2 S_  {19}$ & $Y_  {25}$  & & & \\ 
 $ q_2  \bar q_2   $          &        &        &      &  &  &  &  \\   
 $  q_1  \bar q_2    $        &    $Y_  {23}$ &  $\phi    $ &     &
 $\phi   +\phi^2Y_  {12}$  & $Y_  {25}  +\phi^2  Y_  {11}  +\phi^2 S_  {10}$ &  $Y_  {25}
 +\phi^2    Y_  {11} +\phi^2    S_  {19}$  &  $Y_  {25}
 +\phi^3   Y_  {22} +\phi^4    Y_  {19,20}$  \\   
$ q_2  \bar q_1    $  &      &       &        &   $Y_  {23}$ &  $Y_  {23}$
&  $Y_  {23}$  &  $\phi    +\phi^2 Y_  {21}$  \\ 
\hline   \end{tabular} \end{center}  \end{table}

%%%%%%%%%%%%%%%%%%

\section{Discussion and  conclusions}
\label{sec3}

Let us first state some general features of our results.  The top-down
construction for the $ Z_{6-II}$ orbifold has a large number of flat
directions and a rich structure of superpotential couplings.  This
contrasts with the situation prevailing for the $Z_3$ orbifold
models~\cite{giedt02,giedt05,Font:1989aj}, but is in harmony with that
in non-prime orbifolds or free fermions.  Comparing with the string
spectra obtained in intersecting brane
models~\cite{leontaris01,everett02} would not be very teaching because
the existing type $I$ string models are more akin to bottom-up type
constructions.

We have carried the F flatness test indiscriminately for both types $
A$ and $B$ flat directions, restricting to the superpotential
monomials with at most four distinct singlet field factors, $ W_s
(\phi _i) = \prod _{i=1} ^n \phi ^{s_i}_i , \ [n \leq 4, \ s_i \leq
2]$. These contain a subset of the monomials of absolute order $ \leq
2n = 8$.  It is intuitively clear that the F flatness condition is
more severe for the lower order superpotential monomials, since these
are more likely to contain all (or all but one) of the fields excited
along the flat direction.  Upon pushing the F flatness test in model $
B$ to the order $ n = 5 $ of $W_s$, we find that out of 100 order
$n=4$ flat monomials, 34 are lifted by the 3 monomials of order $ n=3$
in $ W_s= S_ { 1}Y_ {19}^2 , \ Y_ { 4}S_ { 5}Y_ {25} ,\ S_ { 5}Y_ { 5}
Y_ {23}.$ Including all the monomials in $ W_s$ up to order $ 8$,
selects, $0, \ 6,\ 40$ flat monomials of orders $ n=2, \ 3,\ 4$.

To determine which of the singlets decouple and which remain massless,
we have derived the mass matrices for the $ S_i $ and $ Y_i $ by
scanning over each individual flat direction up to absolute order
8. We found that one cannot decouple all the singlets at the string
scale by these means in any of the models.  However, taking the
combinations of all the singlets of a given order, we found that all
singlets of models $A1$ and $A2$ can be decoupled if the order is
large enough, $n \geq 5 $ and $ 7 $ respectively, and the number of
fields involved is itself large enough. One might prefer vacua defined
by smaller sets of fields, and indeed our aim here is not to decouple
all the singlets but leave a few of them massless in order to obtain
an NMSSM like model. We thus prove that there is room for decoupling
all but a few relevant modes, should one perform a thorough scan over
the flat directions.  As far as model $B$ is concerned, the number of
singlet fields is too high to decouple all of them at the order
considered, and making this model viable on a phenomenological basis
would require giving a lot of singlet modes a mass at the level of the
Pati-Salam breaking down to the Standard Model gauge group.

Our study of the fermion mass generation has been rather sketchy.  The
preferred candidate for the electroweak Higgs bidoublet is obvious
only in model $A1$.  To discriminate between the heavy and light
flavors in models $A2 $ and $B $ requires a closer analysis of the
couplings of bidoublets and bifundamentals. Such a task would be
warranted once one has really in hand a benchmark type model.  All
three models satisfy at the string mass scale a $D_4$ family symmetry
with a number of bifundamental modes $ f $ or $ f^c $ from the twisted
sectors $ T_1$ or $T_3$ transforming as doublets.  An acceptable
description of the fermions flavor structure can be achieved only if
the $D_4$ symmetry is broken.  On side of the promising mechanism
using the condensation at the Pati-Salam breaking scale of the
composite singlet fields, $ {\cal O}^c _{ij} = f^c _i\bar f^c _j $,
stringy mechanisms can be envisionned to lift the degeneracy of the
modes in the $ T_{1,3} $ sectors. One could use blow up submanifolds
of different sizes at the two fixed points, or consider the so far
lightly explored models with three Wilson lines~\cite{lebedev08}.

The existence of a light pair of Higgs bosons seems to be correlated
with the presence of massless exotic color triplets descending from
sextet and bifundamental modes, $ C_i, \ f^{'c}_i $.  This is
unavoidable in models $A1$ and $B$ because of certain $ h_i$ and $C_i
$ having identical quantum numbers.  A single pair of $ q_i , \ \bar
q_i $ modes always fails to decouple.  Since the exotic modes are
vector like, they could pair off or couple to lighter modes into which
they decay, through lower scale physics, including the $PS \to SM $
transition.  We have not examined here the decoupling of the weak
doublet and singlet exotic modes, $d ^l = (N ^l, E^l ), \ d ^r = ( E
^r, N^r) $, since these represent perhaps a lesser threat on the low
energy theory.

Our main purpose in this study was to establish an existence proof,
based on three representative string models, of supersymmetric models
where suppressed bilinear couplings of the electroweak Higgs doublets
coexist with unsuppressed trilinear couplings to singlets, $ \mu h_k
h_l + \l h_k h_l \phi , \ [\mu \leq O(\phi ^8), ] $.  Rather than
scanning over the flat directions, we have followed an approximate
procedure making use of what we term as composite flat directions.
This is admittedly open to criticism, especially at high orders,
because of the large number of excited singlets.  However, the
qualitative orientation this approach gives does not seem invalidated
by selecting randomly, or scanning in a global way, a subset of
individual flat directions.  We frequently encounter canonical and
non-canonical type trilinear couplings, $ \l _{ij} ^kh_i h_j \phi _k
$, however, without any obvious correlations between these bidoublets
and those dominating the fermion-Higgs Yukawa couplings.

Additional $U(1)'$ gauge symmetries are generally present in the
individual flat directions of order $ n \leq 4$, but they are absent
in most composite directions. It is difficult to decide if a gaugino
supersymmetry breaking mediation or a secluded sector scenario is
favored since our searches give no clue on the size of the $Z'$ boson
mass scale and the $ Z-Z'$ mixing angle.

In summary, the search of models with extra singlets is considerably
facilitated in the $ Z_{6-II}$ orbifold by the rich mini-landscape of
vacua.  We find a clear preference towards an active role for extra
singlets, but the presence of sizeable trilinear couplings $hh \phi $
along with strongly suppressed bilinear couplings $ hh$ is not
systematic.  The study of solutions was made tractable thanks to the
restriction to low orders, $ n\leq 4$.  Beyond this order one must
consider representative samples of the flat directions.  However, the
composite flat directions of low orders, $ n \leq 4 $, seem to capture
general features not invalidated by global scans of the individual
directions, as confirmed by the similar conclusions for models $A1$
and $B$.  The strong $O(\phi ^8)$ suppression required for the $\mu $
couplings suggests, however, that definitive conclusions could only be
made until the searches include higher order couplings.  Weak and
strong points are present in all three models, with no model faring
best on all issues.

%%%%%%%%%%%%%%%%%%%%%%%%%%%%%%%%%%%%%%%%%%%%%%%%%%%%%%%%%%%%%%%%
%%%%%%%%%%%%%FIN DU TEXTE %%%%%%%%%%%%%%%%%%%%%%%%%%%%%%%%%%%%%%

\section{Acknowledgements}
\label{acknow} 
We   would  like  to thank  Dr. M.~Ratz for advice  and useful  discussions. 
One of the authors~(P.H.)  thanks  also R.~Kappl for useful
discussions  and is grateful  to the Alexander Von Humboldt 
Foundation for  the  financial  support  granted to  him  during the
elaboration of the present paper.

\appendix
\section{Flat directions in heterotic string compactification}
\label{appexa}

Given a  gauge theory   with  the Lie group $ G_a$   and  
superpotential $W$,     which includes  the  set of massless chiral  
supermultiplets, $ \phi _i , \ [i=1, \cdots , N] $ 
then the  D flatness conditions are given by: $ 0=D_a = (\phi ^\star _ i
(T^a)_{ij} \phi _ j ) $, where $ T_a$  are the Lie algebra generators,   
and the F flatness conditions are given by: $ W_i\equiv {\dh W \over \dh \phi _i } =0$.
 In the   scalar  fields $ \phi _i $  vector space, 
the D flat directions are parameterized by the gauge
invariant monomials, $ P _\a (\phi _i ) $,
satisfying the equations, $ P_{\a , i} \equiv  {\dh P _\a \over \dh
  \phi _i }   = c _\a \phi _i^\star , $   for suitably chosen complex
constants, $c _\a $,   with  the  $\phi _i$  substituted  by  their (constant) 
VEV~\cite{Font:1989aj,lutytaylor96}.   (This statement
is easily checked by writing the $G_a$ gauge invariance condition as,
$0= \d _a P _\a = \sum _i {\dh P _\a \over \dh \phi _i } \d _a \phi _i
= c _\a \sum _i \phi _i^\star (T_a \phi ) _i $.)  For an anomalous gauge
symmetry~\cite{dine87,atick88,casas89}  $U(1)_A$  
with the $\phi _i $ charges, $ Q_A ^i = Q_A ( \phi
_i ) $,  and  a finite   charge  anomaly  implying  the presence of a 
Fayet-Iliopoulos term with coupling constant $\xi _A$, the D-flatness
condition is modified to 
\be 
0= D_A =\sum _i \phi ^\star _ i Q^i
_{A} \phi _i + \xi _A ,\ \left[\xi _A = e^{2\phi } M_\star ^2 \d _{GS} ^A =
{g_s^2 M_\star ^2 \over 192 \pi ^2 } {1\over \sqrt {k_A}} Tr ( Q_{A}
), \  k_A = 2 \sum _I (Q_A ^{I})^2  \right]  \label{eqgs1} 
\ee
where  $g_s  = <e^\phi >$ is  the string coupling constant,  
with $\phi $ the 10-d dilaton field, and $Q_A ^I, \ [I=1, \cdots
, 16] $ are the orthogonal basis components of the anomalous charge
generator in the $E_8\times E_8$ group weight lattice.  
For     the   set of   gauge group  factors, 
$ G= (\prod _a G_a) \times U(1)_A $, the  D  flatness conditions      
is  solved  by considering holomorphic invariant
monomials,  $ P _\a (\phi _i ) $,  uncharged with respect to the anomaly
free gauge factors $G_a$, but with finite  anomalous  $U(1)_A$ charge, $ Q_A
(P_\alpha )  \ne  0$,  of opposite sign to $ \xi _A \propto Trace ( Q_{A} ) $.  
In the supergravity context, the F flatness equations  are, 
$ W=0,\   F_i= \dh W / \dh {\phi _i} = 0$, where $ W$
is built from holomorphic monomials invariant under the complete set
of non-Abelian and Abelian  gauge  symmetries,  including the 
global type string  theory symmetries. 
 
Most of our   applications   deal  with the field space  of 
non-Abelian singlets charged under  the Abelian gauge groups,   $ G= (\prod
_{a=1}^{N_a}   U(1)_a ) \times U(1)_A $.    The D-flat
directions are parameterized by the monomials, $ P_\a = \prod _i \phi
_i ^ {r ^\a _i} $, solving the simultaneous equations: $ Q_a (P_\a ) =
0,\ Q_A (P_\a ) \ne 0 $ of sign opposite to $\xi _A $.  By contrast,
the superpotential couplings are constructed from the gauge invariant
monomials with respect to both non-Abelian and Abelian (anomaly free
and anomalous) gauge groups  which  obey  the  string selection rules,
listed   for the orbifold  $ Z_{6-II}$   in Eqs.~(\ref{eqsel}).   
The submanifold of vacua for non-Abelian singlets~\cite{cleav198} is
parameterized by the holomorphic invariant monomials, $ P_\a = \prod
_i\phi _i^{r_i^ \a } , \ [r_i^ \a \geq 0 ]$ described by the
column vectors, $\vec  r^\a=(r^\a_1, \cdots , r^\a_N) \in Z ^N _+
$, subject to the equations, \bea && Q_a (P_\a ) = \sum _i {\dh P _\a
\over \dh \phi _i} (Q_a \phi )_i =P_\a  (\phi ) \times \sum _i r_i^ \a Q_a ^i =0, \
Q_A (P_\a )
%= \sum _i {\dh P _\a \over \dh \phi _i} (Q_A \phi )_i   
= P_\a  (\phi ) \times  \sum _i r_i^ \a Q_A^i = - c_A \xi _A , \ [c_A >0]. \eea These
equations  are solved by  the  simple solution,  $ \vert \phi _i \vert / \sqrt {r^\a
_i}= \vert \phi \vert = 1/c_A $.  The flat
directions  may have $0,\ 1 $ or more dimensions, where the 0-d
case involves a fixed common VEV, $\phi = ( { -\xi _A / \sum _i Q_A^i
r^\a _i } ) ^\ud $, and the 1-d or  higher dimension  cases involve a
single or more  free complex parameters.   When the system of linear equations
is underdetermined, the solutions depend on   free
continuous parameters   whose number identifies with the moduli space dimension, $
\cald \leq N- rank (A) $.  The  vector space of     column  vectors,
$\vec r $,  is generated by the
basis of column vectors, $\vec  r ^ A =(r_1 ^ A , \cdots , r_N^A ) $,
associated to the holomorphic invariant monomials, $ M_A = \prod _i
\phi _i ^{r_i ^A }, \ [A =1, \cdots , \cald ]$ such that any
solution can be written as, $ P^n = \prod _\a M_A ^{n_A } , $ for $
n \geq 1 $ and $ n _A \in Z $ (positive or negative signs).
More conveniently, one can  also consider the superbasis of one-dimensional
invariant monomials, $ P_\a $,  satisfying the condition that
they cannot be factorized into products of two or more  invariant monomials.

The D flat direction described by the monomial solution, $ P _\a (\phi
) = \prod _i \phi_i ^{r_\a ^i } $, can be lifted by F-terms of type
$A, \ W_s^A = (\prod ' _{i \in P_\a } \phi _i ^{ r_\a ^i } )^n $ or
type $B, \ W_s^B= \Phi (\prod ' _{i \in P_\a } \phi _i)^n , \ [\Phi
\not \in P_\a ] $ consisting of gauge invariant superpotential
monomials with all, or all but one, field factors included in $ P _\a
(\phi )$.  In the terminology of~\cite{cleav298,cleaver07,cleav398},
these refer to the gauge invariant non-renormalizable couplings in the
field theory action, prior to applying the string selection rules.
For the F flatness to remain valid to arbitrary orders of the full
superpotential, an infinite number of conditions must be imposed for a
type $A$ direction but a finite number for a type $B$ direction.  The
F flatness conditions are, of course, more restrictive when set on the string
theory action.   

\section{Review of  string construction  for $ Z_{6-II}$  orbifold}
\label{appexb}

The orbifold $ Z_{6-II}$ is described by the rotation vector, $ \T ^I
= e^{2i \pi v_6 ^I } $,  or  the twist vector, $ v_6^I ={1\over 6} (1,2,-3)$,
acting on the  complexified coordinate and fermion field  components 
$ X^I, \ \psi ^I $    associated to the complex planes of $ T^2_I$.
This orbifold is equivalent to the direct
product of suborbifolds, $ Z_2 \times Z _3 $, with twist vectors, $
v_2^I = 3 v_6^I= {1\over 2}(1,0, -1) , \ v_3^I = 2 v_6^I= {1\over 3}
(1, -1,0)$.  The anisotropic compactifications~\cite{Kobayashi:2004ya,Kobayashi:2004ud}
are realized with 6-d tori which factorize
on the three maximal tori $ T^2_I$ of the rank $2$ semi-simple groups,
$ G_2, \ SU(3), \ SO(4) $, with lattice basis vectors $e_{1,2} ,\
e_{3,4} ,\ e_{5,6} $, and orbifold point  group action represented  by  
the $  2\times 2 $ matrices   for the Coxeter   operators  
$ C ^T_I \simeq \T ^I$  in  the lattice bases, 
\bea && \T (G_2) =\pmatrix{1 & 3 \cr -1 & -2 }
 ,\ \T ( SU(3) ) =\pmatrix{0& 1 \cr -1 & -1 } ,\ \T ( SO(4) )
 =\pmatrix{-1& 0 \cr 0 & -1} . \eea

\vskip 0.5 cm \squeezetable \begin{table} \begin{center}
\caption{ \label{taba1} 
Geometric data for the orbifold $ Z_{6-II}$ on the torus
$T^6= T^2_1 \times  T^2_2 \times  T^2_3$ of   lattice,  
$ \L (G_2) + \L ( SU(3) ) + \L ( SO(4) ) $.  The line entries   refer
  to the  twisted sectors, $ T_g , \ [g=1, \cdots , 4] $  of $ Z_{6-II}$  
  with $ T_1 \sim T_5$,      and the  associated $T_{g_1,
  g_2} $ sectors of the orbifold, $ Z_2 \times Z _3 $.  The second
  colum displays the number of fixed points $ \caln _{fp} ^I $ in the
  three complex planes $ T^2_I $.  The third column consists of three
  subcolumns which display for the lattice $ \L (G_2)$, the fixed
  points and shift vectors, $ f_a, \ u_a $ and $f_{ab}, \ u_{ab} $, 
for the    elements $\T ^{2,4} $ and $\T ^{3} $,   
the linear combination eigenvectors, $ \vert \g (g) > $, and  
the eigenvalues $\g (g) $,  with $\o = e^{2i\pi /3} $.   The fourth column consists of two
  subcolumns which display for the lattice $ \L (SU (3) )$, the fixed
  points and shift vectors, $f _{n_3}, \ u_{n_3} $,  for the
  elements $\T ^{1} $ and $\T ^{2,4} $,  and the range of
  $n_3$.  The fifth column consists of three subcolumns which display
  for the lattice $ \L (SO (4) )$,  the fixed points and shift vectors,
  $ f _{n_2, n'_2}, \ u_{n_2, n'_2} $,  
for the   elements $\T ^{1} $ and $\T ^{3} $,
and the range of $n_2, \ n'_2$.   The last column displays the modes
multiplicities   $\cddd $      assigned in the  various  twisted
sectors $T_g$,  in the  case $ W'_2=0$   with  unresolved quantum
number $n'_2 =0, \  1$  for the orbifold   group action on $ T^2_1$. 
The  entries for $\cddd $  correspond to  the entries  for the   
eigenvectors, $ \vert \g (g) > $, of eigenvalues $\g (g) $.} 
\vskip 0.5 cm \begin{tabular}{|c|c| ccc||cc||ccc||c|} \hline
${Z_{6-II} \atop Z_2 \times Z_3 }$ & $ \caln _{fp} ^{I} $ & $ T^2_1 =
\L (G_2) $ & $ \vert \g (g) > $ & $ \g (g) $ & $ T^2_2 = \L ( SU(3) )
$ & $ n_3 $ & $ T^2_3 = \L ( SO(4) ) $ & $ n_2 $ & $n'_2 $ & $\cald $
\\ \hline &&&&&&&&&& \\ $ {T_{1} + T_{5} \atop T_{1,2} + T_{1,1} } $ &
1,3,4 & $ {f_\a = 0\atop u _\a =0 } $ & 1 & 1& $ { f _{n_3} = { [ 0, \
{2 \over 3} e_3 + {1\over 3} e_4 \atop {1 \over 3} e_3 + {2\over 3}
e_4 ] }   \atop u _{n_3} = 0, e_3, e_3 +e_4 } $ & $ 0,1,2$ & $ {f
_{n_2, n'_2} = {n_2 \over 2}e_5 +{n'_2 \over 2} e_6 \atop u _{n_2,
n'_2} = -n_2 e_5 -n'_2 e_6 } $ & $ 0,1 $ & $ 0,1 $ & 2 \\ &&&&&&&&&&
\\ $ {T_{2} +T_{4} \atop T_{0,1} + T_{0,2} }$ & $ 3,3 ,1 $ & $
{f_a={a\over 3} e_1 , \ u_a= -ae_2 \atop [a=0,1,2]} $ & ${f_0,{1\over
\sqrt 2 } (f_1+f_2) \choose {1\over \sqrt 2 } (f_1-f_2) } $ &$
{1\choose -1} $ & $ f _{n_3} , \ u_{n_3} $ & 0,1,2 & $ {f _\a =0 \atop
u _\a =0} $ & $0$ & $ 0 $ & $ {2\choose 1}$ \\ &&&&&&&&&& \\ $ {T_3
\atop T_{1,0} } $ & 4,1,4 & $ {f_{ab}= \ud (ae_1+be_2) \atop {
u_{ab}=ae_1+be_2 \atop [a,b=0,1] } } $ & $ \pmatrix{f_{00}, \ {f_{01}
+ f_{10}+f_{11} \over \sqrt 3} \cr {f_{01} +\o f_{01}+ \o ^2 f_{11}
\over \sqrt 3 } \cr {f_{10} +\o ^2 f_{01}+ \o f_{11} \over \sqrt 3 }
} $ & $ \pmatrix{1\cr \o ^2 \cr \o } $ & $ {f _\a =0 \atop u_{\a } =0
} $ & 0 & ${f _{n_2, n'_2} \atop u _{n_2, n'_2} }$ & $ 0,1 $ & $ 0,1 $
& $ \pmatrix{4 \cr 2 \cr 2 } $ \\&&&&&&&&&& \\ \hline
\end{tabular} \end{center}   \end{table} \vskip 0.5 cm

The input data for the orbifold fixed points and shift vectors in the
various twisted sectors is displayed in Table~\ref{taba1}.
%\bea && T^2_1 \sim \L ( G_2 ):\ f_a= {a\over 3} e_1 ,\ u_{ f_a}= -a
%e_2 , \ [ a=0,1,2 ]; \ f_{a,b} = {a\over 2} e_1 + {b\over 2} e_2 ,\
%u_{ f_{a,b} }= ae_1 + be_2 ,\ [a , b = (0,1) ] . \cr && T^2_2 \sim \L
%( SU(3) ):\ f _{n_3} = [0, {2 e_3 + e_4\over 3 } , \ {e_3 + 2 e_4
%\over 3 } ] ,\ u _{n_3} = ae_3+be_4, \ [ n_3 = a+b  \  \text{mod} \ 3 ] .\cr &&
%T^2_3 \sim \L ( SO(4) ):\ f _{n_2, n'_2} = \ud (n_2 e_5 + n'_2 e_6 ),\
%u_{n_2, n'_2} = - n_2 e_5 - n'_2 e_6 .  \eea 
The shift vectors $u _{g,f}\in \L $, associated to the fixed points $f$
in sectors $T_g \sim \T ^g $, are defined  by, $ (1 -\T ^g ) f = u
_{g,f} $,  modulo   elements of the    sublattice,   $ \L _{\T ^g} =
(1-\T ^g ) \L $,   of   the torus lattice $ \L $. The  shift vectors   
are  in  1-to-1 correspondence with the  orbifold
space  group  elements,  $( \T ^g, u _{g,f} )$, which act on the
orthogonal  frame coordinates as,  $ X \to    \T ^g  X +   u _{g,f} $. Each shift  vector $u
_{g,f}$   represents  a  conjugacy class  element  of the  torus
lattice  coset, $     \L /   \L _{\T ^g}  .$  
The fixed points $f$, shift vectors $u_f$ and sublattices $\L _{\T ^g } ^I=
(1-\T ^{g} ) \L (T_I^2) $ of the twisted sectors $g$ are defined as
follows for the three 2-d tori, $T_I^2 $, with $ G_2 , \ SU(3) , \
SO(4) $ group weight lattices: 
\bea && \L (G_2):\   (1-\T ^{2,4} ) f _{a} = u
_{a} = -a e_2 , \ f_a= {a\over 3} e_1 ,  \ [ a=0,1,2 ] ; \
(1-\T ^{3} ) f_{a, b} =u_{a, b}  = {a\over 2} e_1 + {b\over 2} e_2 ,\ 
[a , b = (0,1) ] ; \cr && \L ^1_{\T } \sim ( e_1, e_2)
,\  \  \L ^1 _{\T ^{2,4} } \sim ( e_1, 3e_2) ,\ \ \L ^1_{\T ^{3} }  \sim
(2e_1,2e_2)  . \cr && 
\L (SU(3)):\ (1-\T ^g ) f _{n_3} = u _{n_3} = [e_3,
e_3 + e_4]= n_3 e_3 , \  f _{n_3} =
[0, {2 e_3 + e_4\over 3 } , \ {e_3 + 2 e_4 \over 3 } ] ,\ 
[n_3 = 0,1,2  ] ;  \cr && \L ^2_{\T ^g } \sim ( e_3-e_4, 3e_3)  , \
[g= 1,2,4] . \cr && 
\L (SO(4)):\ (1-\T ^g ) f _{n_2,
n'_2} = u _{n_2, n'_2} = -n_2 e_5 - n_2' e_6 ,\ 
f _{n_2, n'_2} = \ud (n_2 e_5 + n'_2 e_6 ), \ [n_2, \ n'_2 = 0,1 ];  
\cr && \L ^3_{\T ^{g} } \sim (2e_5 ,2e_6)  , \ [g=1, 3]. 
\label{eqlatt3}\eea

%  These    are  defined for the $ SU(3) $ group weight lattice  by,
%  $ (1-\T ^g ) f _{n_3} = u _{n_3} =   a  e_3 +  b e_4   \simeq n_3 e_3 , \ [
% n_3 = a+b  \  \text{mod} \ 3 = 0,1,2 ; \ g= 1,
% 2, 4]$ and for the $ SO(4) $ group weight lattice by, $ (1-\T ^g
% ) f _{n_2, n'_2} = u _{n_2, n'_2} = -n_2 e_5 - n_2' e_6 ,\ [n_2, \
% n'_2 = 0,1 ; \ g=1, 3 ] $. 

The orbifold fundamental group  allows for
three discrete Wilson lines:  $ W_2, \ W'_2$ of order $ N_2= N'_2=2$,
around two dual one-cycles of $T^2 _3$,  and $ W_3$ of order $ N_3= 3 $
around the $e_3$ one-cycle of torus $T_2 ^2$. Upon turning on the Wilson lines, the $T_g$
twisted sectors split into twisted subsectors, $ (g , \g , n_3 , n_2,
n'_2) $, labelled by the discrete parameters, $ n_3 =0, 1,2 $ and $n_2
= 0, 1,\ n'_2 = 0, 1 $, for the lattices $\L ( T^2_2) \sim \L (SU(3)) $ and
$ \L (T^2 _3) \sim \L (SO(4)) $. The  complex phase parameters,  $\g
(g)$, describing   the action of $\T ^g$ on the fixed points of the  
lattice $\L ( T^2_1) \sim \L (G_2)$  are needed to specify the
degeneracy of string modes  in the sectors $g=2,4 $ and $g=3$.  
b
The string states are described by the $SO(8)$ group weight vectors, $
r^a , \ [a=1, \cdots , 4]$ for the right-moving fermion fields; the $
E_8 \times E_8$ group weight vectors, $P^I , \ [I=1, \cdots , 16]$ for
the left-moving  16 coordinate fields; and the oscillator
numbers, $ N^R _i, \ N^L _i \in Z , \ [i=I, \bar I; \ I, \ \bar I
=1,2,3]$ for the 3 right- and left-moving complex coordinate fields $
X^I$ of $T^6$.  The string squared mass spectrum is evaluated in terms
of these quantum numbers  by means of  the formula  \bea &&
{M ^2 _R \over 8 m_s ^2 } = \sum _I ( N_I^R \o _I ^{(g)} + N^R_{\bar
I} \o _{\bar I} ^{(g)} ) + \ud \sum _{a=1}^4 (r ^a +g v ^a ) ^2 - E_0
^{(g)} - \ud , \cr && {M ^2 _L\over 8 m_s ^2 } = \sum _I (N_I^L \o _I
^{(g)} + N^L_{\bar I} \o _{\bar I} ^{(g)} ) +\ud \sum _{I=1}^{16} (P^I
+ X ^I_{g,n_f} ) ^2 - E_0 ^{(g)} - 1 , \cr && [E_0 ^{(g)} = \ud \sum
_I \vert \hat {gv_I} \vert( 1 - \vert \hat {gv_I} \vert ) , \ \hat
{gv_I} = {gv_I} \ \text{mod} \ 1 = {gv_I} - [gv_I] , \cr && X_{g, n_f
}= g V + n_{f, a} W_a = gV + n_{3} W_3+ n_{2} W_2+ n_{2} ' W'_2 ]
\label{eqms2} \eea 
(in units of $ m_s$) where   $X_{g, n_f } $ denote
the shift vectors in the $ E_8 \times E_8$  group lattice  depending on the orbifold
and Wilson lines  gauge  embeddings, $ V, \   W_3, \  W_2 , \ W'_2$. 
The oscillator energies are defined by $ \o _I ^{(g)} = \hat {gv_I} $ for
$ \hat {gv_I} > 0$ and $1 -\vert \hat {gv_I} \vert $ for $ \hat {gv_I}
\leq 0$, with $ N_{I, \ \bar I } ^{L,R} \in Z _+ $. An equivalent
definition is, $ \o _I ^{(g)} = gv_I \ \text{mod} \ 1 , \ \o _{\bar I}
^{(g)} = - gv_I \ \text{mod} \ 1 ,$   for the determinations  obeying,
$0 < \o _{I, \bar I} ^{(g)} \leq 1 $. In  the $Z_{6-II}$ orbifold, the string states  
generally occur  in  CPT conjugate pairs of opposite 4-d chirality  for
the sectors, $ g $ and $ g'=6-g $,   except for
the twisted sectors $ T_2 $ and $T_4$ with $ g=2, \ 4$,  which admit
both left and right chirality modes each (because of the supersymmetry
$\caln =2$), and the (self-conjugate) twisted sector $T_3$ with $ g=3
$.  The vector and spinor weight vectors $ r^a_{v,s} $  of the $SO(8)$
symmetry group   for  right moving    fermions   are associated to
boson and fermion superpartners of the massless chiral supermultiplets  
as  displayed in Table~\ref{taba2}.

\vskip 0.5 cm %\squeezetable
\begin{table} \begin{center}  
\caption{ \label{taba2} 
The $SO(8) $  group weight vectors,  $r^{l,r}_v $ and $ r^{l,r}_s$,  
assigned   to the massless right moving  modes, $ M_R ^2 =0$,  with  left and right
chiralities  ($l, \ r$). The vector and spinor representations,
($v, \ s $)   differ by the  weight  vectors $\rho _{l, r} $  assigned
to  the supercharge generators, $r^{l,r}_s= r^{l,r}_v + \rho _{l, r} , \  [\rho _{l, r} =(\pm   \ud
\pm \ud  \pm \ud , \pm \ud )  ] $.       The line  entries refer to  the
untwisted and twisted sectors,  $ g= 0, 1 , \cdots , 4 $. 
No massless mode  solutions   arise  for right movers
carrying oscillator exitations, $ N ^R_{I, \  \bar I} $.}
\vskip 0.5 cm \begin{tabular}{|ccc||ccc|} \hline $ g $ & $ r^l_v $ & $ r_s ^l = r_v
+\rho _L $ & $g$ & $ r^r_v $ & $ r_s ^r = r_v +\rho _R$ \\ \hline 
$ 0 $ & $ (\underline{100},0) $ & $ (\underline{+\ud -\ud -\ud
},-\ud ) $ & $ 0 $ & $ (\underline{-100} ,0) $ & $ (\underline{-\ud
+\ud +\ud },+\ud ) $ \\
$ 1 $ & $ (001,0) $ & $ (-\ud -\ud +\ud ,-\ud ) $ & $ 5 $ & $(-1-2 2
,0) $ & $ (-\ud -\td +\cd ,+\ud ) $ \\
$ 2 $ & $(001,0) $ & $ (-\ud -\ud +\ud ,-\ud ) $ & $ 2 $ & $ (-1-11,0)
$ & $ (-\ud - \ud \td ,+\ud ) $ \\
$ 3 $ & $ (0-12 ,0) $ & $ (-\ud -\td +\td , -\ud ) $ & $ 3 $ & $
(-1-11,0) $ & $ (-\ud -\ud \td ,+\ud ) $ \\
$ 4 $ & $ (0-1 2 ,0) $ & $ (-\ud -\td + \td, -\ud ) $ & $ 4 $ & $
(-1-2 2 ,0) $ & $ (- \ud -\td +\cd ,+\ud ) $ \\ \hline
\end{tabular} 
\end{center}    \end{table} \vskip 0.5 cm

The GSO projection on the orbifold singlet states is determined 
via the one-loop partition function by the condition, 
$P (g, n_f, \g , \phi ) =1, $     where 
\bea && P (g, n_f, \g , \phi ) = {1 \over N} \sum
_{h=0} ^{ N-1} \D ^ h (g, n_f, \g , \phi ) , \cr && [\D ^ h (g, n_f,
\g , \phi )= \g (g,h) \phi (g,h) e ^{ 2 i \pi [ (P + X _{g, n_f } )
\cdot X _{h, n_f }- (r + g v )\cdot h v - \ud ( X _{g, n_f } \cdot X
_{h, n_f } - gh v^2 ) ] } ] .
%\cr &&  X_{g, n_f }= g V + n_{f, a} W_a   = gV +n_{ f, 3} W_3+ 
%n_{f, 2} W_2+ n_{f, 2} ' W'_2] ,  
%\ [\phi = e ^{ 2 i \pi \sum _{i= I, \bar I } ( N_i ^L - N_i ^R ) \hat
%\phi _I } ] . 
\label{eqproj} \eea 
The twists along the world sheet spatial and temporal directions are
denoted by $g $ and $h $ and the terms $ \D ^h $ include the complex
phases,  $\g (g, h) = \g ^h (g)   $,  eigenvalues of the orbifold twist action on the
fixed points in the 2-d torus lattice $ \L (T_1 ^2 )= \L (G_2) $, and
the complex phases from oscillator excitations,  $ \phi (g, h) =
\phi ^h (g)  $,   
\bea && \phi (g) = e ^{2 i \pi \sum _ {i= I, \bar
I } ( N_i ^L - N_i ^R ) \hat \phi_i }, \ [\hat \phi _I = v_I \text{sgn} (
\hat {gv_I} ) , \ \hat \phi _{\bar I } = - v_I \text{sgn} ( {\hat {gv_I} } ) , \
I, \bar I =1, \cdots , 3] .  \label{eqgso1} \eea
%\phi_L = e^{2 i \pi (N_I^L - N _{\bar I}^L \text{sgn} ( {\hat {gv_I} } ) )
%\vert v_I \vert }, \ I, \bar I =1,2,3]
%  The terms $ \D ^ {h=0} $ give the multiplicities of string modes.  
The level matching for right and left movers entails the conditions, $
N ( X _{g, n_f }^2 - (g v )^2 ) ) = 0 \ \text{mod} \ 2 $,  with
additional conditions involving the scalar products of Wilson lines, $ W_3,
\ W_2, \ W'_2.$ For the twisted sectors, $ T_{2, 4}$ and $ T_3$, 
the projections for the $Z_3$ and $ Z_2$
suborbifolds     can be implemented by summing over the time twists in the subsectors $ (h=3,
\ m_2,\ m'_2 )$ and $ (h=2, \ m_3) $ as follows 
\bea && \bullet \ g= (2, \ 4 ),
\ h=3 :\ P (g, n_3, \g , \phi ) = \g ^h (g) \phi ^h (g)  {1 \over 4} \sum
_{m_2, m'_2=0,1} e ^{ 2 i \pi [ (P + \ud (g V + n_3 W_3 ) ) \cdot (h
V+ m_2 W_2+ m'_2 W'_2 ) - (r + \ud g v )\cdot h v ]} , \label{eqgso2}
\cr && \bullet \ g=3, \ h=2:\ P (g =3, n_2, n'_2, \g , \phi ) = \g ^h (g) \phi
^h (g) 
{1 \over 3} \sum _{m_3=0,1,2} e ^{ 2 i \pi [ (P + \ud (g V + n_2
W_2 + n'_2 W'_2 ) ) \cdot (h V+ m_3 W_3 ) - (r + \ud g v )\cdot h v ]}
 \cr && \label{eqgso3} \eea 
Except for the   singly twisted sector, $T_1$,   where 
the conditions in Eq.~(\ref{eqproj})   hold with fixed values of $ n_3,
\ n_2$,   the projections in the
other  sectors can  be    concisely stated in terms of   simple  sets of
 conditions  on the  weight vectors $ P ^I$.  For the untwisted
sector, these conditions are: $ P \cdot V \in Z, \ P \cdot W_2 \in Z, 
\ P \cdot W'_2 \in Z, \ P \cdot W_3 \in Z.$  For the twisted sectors, $ T_{2, 4}$ and $ T_3$, the
projections  are  more easily  implemented by imposing the conditions \bea &&
\bullet \ g= (2, \ 4), \ h=3 :\ {\D ^h =(\g \phi ) ^h e ^{ 2 i \pi h
[(P + \ud (g V + n_3 W_3) ) \cdot (V + n_3 W_3) - (r+\ud gv) \cdot v ]
} =1 \choose h \tilde P \cdot W_2 \in Z , \ h \tilde P \cdot W'_2 \in
Z , \ [\tilde P = P + X _{g, n_3,0,0} ] } .  \cr && \cr && \bullet \ g=3, \
h=2:\ { \D ^h = (\g \phi ) ^h e ^{ 2 i \pi h [(P + \ud (g V + n_2
W_2 + n_2' W'_2 ) ) \cdot (V + n_2 W_2 + n_2' W'_2 ) - (r+ \ud gv)
\cdot v ] } =1 \choose h \hat P \cdot W_3 \in Z , \ [\hat P = P +
X_{g,0,n_2,n'_2 } ] } .
\label{eqgso3p}  \eea   
This is the  prescription for the GSO orbifold phase  that we have
used to obtain the  string spectra  of models $A1, \ A1, \ B.$ 
Use of the  world sheet  modular invariance  (level matching
conditions)    allows replacing  the conditions on the scalar
products in Eqs.~(\ref{eqgso3p}) by  the simpler ones~\cite{Kobayashi:2004ya}: $P\cdot W_2 \in
Z , \ P\cdot W'_2 \in Z  $  for $ g= (2, \ 4 )  $ and   $  P \cdot W_3
\in Z $  for $ g=3$.   The following  alternative 
projections  are also proposed in~\cite{Buchmuller:2005jr} \bea
&& \bullet \ g= (2, \ 4 ) :\ { 3[(N^L _i - N ^R _i) \hat v_i + q_\g
  (g)   - (r+gv) \cdot v + (P+gV + n_3 W_3) \cdot V ] \in Z \choose (P+gV + n_3
W_3) \cdot {W_2 \choose W'_2 } \in Z } , \label{eqgso4} \cr && \cr && \bullet
\ g=3:\ { 2[(N^L _i - N ^R _i) \hat v_i + q_\g (g) - (r+gv) \cdot v +
(P+gV + n_2 W_2 + n'_2 W'_2 ) \cdot V ] \in Z \choose (P+gV + n_2 W_2
+ n'_2 W'_2 ) \cdot W_3 \in Z } ,\label{eqgso5} \eea where $ \g (g) =
e^{2i\pi q_\g (g) } $.  It is important to remark at this point that the
orbifold projection for sectors with fixed $ T^2$ subtori boils down
to the  invariance under the world sheet   modular group.  The modular   invariance  is vital if the mixed
gauge and gravitational anomalies in the various Abelian gauge factors
$ U(1)_a$ are to satisfy the universality relations~\cite{dine87,kobaya97} 
  \bea && {1\over 24 k_a ^ {1/2} } Tr(Q_a)
= {1\over 3 k_a ^{3/2} } Tr(Q_a^3) = {1\over 2 k_a ^ {1/2}  } Tr(Q_a
c_b(R) ) = {1\over k_a ^ {1/2} k_b } Tr(Q_a Q_b^2 ) = 8\pi ^2\d _{GS} ^a
,\label{equniv} \eea 
which   are  necessary for the Green-Schwarz type anomaly
cancellation. mechanism   to  work. When the  orbifold space group action   by  $ h = (\T ^h, u
_{h,f } ) $  along the string time direction  does not commute with
that of the constructing orbifold  element  along the   string space
direction,  $g= (\T ^g , u _{g,f } )$, the physical
states must be constructed by summing over elements of the orbit, $h
^n g h^{-n} $, and including suitable complex  phase factors to ensure
the  $SL(2,Z)$ modular group
invariance~\cite{Lebedev:2007hv,Lebedev:2006kn,deGroot08}.  
This  situation  indeed   occurs for the
sectors, $ g=2, 4$ and $g=3$, where the fixed points in $T^2 _1$ with
lattice $\L (G_2) $ are reshuffled by the action of $h$. The
physical string modes decompose  then into eigenvectors of $\T ^g ,\
[g=2,3,4] $ with eigenvalues  $ \g (g) =\pm 1 $ in the ($ g=2,\ 4$)
sectors $ T_2 ,\ T_4 $ and $ \g (g ) = 1, \o , \o ^2 , \ [\o = e
^{2i\pi /3} ] $ in the ($g=3$) sector $ T_3 $.   
The degeneracy of eigenstates is $\cald = 2$ in $T_1$; $\cald =
(1, 2) $ for $ \g = (1, -1) $ in $T_2,\ T_4$,  and    $\cald = (4, 2, 2
) $ for $ \g = ( 1, \o ^2, \o ) $ in $T_3$,  as displayed in
Table~\ref{taba1}. 

For  non-commuting twists, $ [h,g]\ne 0$,   the complex phase $ \g (g , h) $ is not
determined in a unique way by the GSO projection.     Since $ \g (g, h )
$ fixes the degeneracies of string modes, its choice    has  clearly an incidence
on the   $ U(1)_a$  groups anomalies   and hence on the  modular
invariance.   The choice of  $ \g (g, h )$   can be  set uniquely by requiring the mixed 
gauge and gravitational anomalies to
satisfy the  anomalies universality relations in Eq.~(\ref{equniv}).  Given a
massless string state, specified by the quantum numbers, $ P ^I , \
R^I ,\  N_i ^L ,$ then the  freedom on $\g (g, h  ) $ is taken into account by
redefining, $\g (g, h ) \to   \chi (h) \g (g, h ) $, 
in terms of  the  GSO modified condition, $ \D ^{h}
\chi ^{h}= 1 $,  so  that the  mode multiplicities  determined by 
$\g (g,h) $ give    anomaly coefficients  satisfying the  requisite  universality  relations. 
Although the natural choice, $\chi =1$, applies   for 
the large  majority of modes,   there do occur cases where
a non-trivial $\chi $  in $ \D ^h $ is needed  to satisfy the universality
relations.  The     factors $ \chi (h)$   depend  on the 
prescription used  for the modified gamma phases. 
With our  GSO projection  prescrition  in Eqs.(\ref{eqgso3p}),  
one exception     occurs  for  three modes of sector $ T_3 $ in model $A2$  
and  another one  for   two modes  of   sectors $ T_2 $ and $ T_4 $   in model $B$. 
%by setting   $\chi  = \o $ (by  inserting the factor,  $ \chi  = -1$).  

Finally, we quote the string selection rules on the $n$-point world
sheet correlators of vertex operators, $ < \prod _{l=1} ^n V _l ( z_l,
\bar z_l ) > $. Applied  to  the superpotential couplings, the
conditions   set  by the gauge symmetries, the $H$-momentum conservation, and the orbifold
point and space symmetry groups  read: \bea && \bullet \ \sum _{l=1}
^n [P ^I
(l) + X  ^I _{g, n_f} (l) ] \in \L (E_8 \times E_8) , \ [X_{g,f} = g V +
n_{g,f} ^a W_a ] . \cr && \bullet \ [\sum _{l=1} ^n R_I(l) -1 ] = 0 \
\text{mod} \ ({1 \over \vert v_I\vert } ) = 0 \ \text{mod} \ (6, 3, 2)
, \ [R^I(l) = r ^I (l) +g_l v ^I (l) - N ^{L}_I (l) + N_{\bar I} ^L
(l) ]  . \cr && \bullet \ \sum _{l=1} ^n \T _I^{ g_l} = 0\ \text{mod} \ 1
\ \Longrightarrow \ \sum _{l=1} ^n g_l = 0\ \text{mod} \ N  . \cr && \bullet \
\sum _{l=1} ^n (1 - \T _I ^{ g_l} ) ( f _l - \L _I ) = 0\ \text{mod} \
N \ \Longrightarrow \ \sum _{l=1} ^n n_3 (l) = 0 \ \text{mod} \ 3 , \
\sum _{l=1} ^n ( n_2 (l) , n_2 '(l) ) = ( 0 , 0) \ \text{mod} \ 2 .
\label{eqsel} \eea 
Some of the above selection rules  can be formulated as ordinary~\cite{Kobayashi:2006wq,ko07}
or $R$ type~\cite{araki07}  discrete symmetries. 
We have omitted from the list in~Eq.(\ref{eqsel}) the constraints on 
couplings  from  the space orbifold group action on  the $T^2_1$ torus with lattice
$\Lambda (G_2)$. These consist of the following two selection rules,
stated around Eq.~(C.9) of~\cite{Kobayashi:2004ya}:
\begin{itemize}
\item Rule $I$ on the gamma phases of physical modes: $\prod _{l=1}^n
\gamma _l =1$.  Using the definition, $\gamma _l \equiv  \g (g_l )= e^{2i\pi q
_{\gamma _l}} $, this can also be written as, $\sum _{l=1}^n q
_{\gamma _l} = 0 \ \text{mod} \  1 $.  \item Rule $II$ for 
couplings involving the twisted sector modes $T_{2,4 } $ only, or
twisted sector modes $T_{3} $ only, times untwisted  sector states.   
The version of this rule derived in~\cite{Buchmuller:2006ik}   states
that the column vector of gamma phases for fields in  $n$-point couplings  
 $ T_{2,4} ^{p_1} T_0^{p_2}  $ or $ T_{3} ^{p_1}T_0^{p_2} $, with $
p_1\ne 0,\ n>  p_2 >0 $   must  obey the  condition, $
\vec q_{\gamma _l }  \not \in [ (p, 0, \cdots , 0) + \text{perms} ], \ [p
\ne 0 ] $. Thus, the allowed couplings are those containing 
at least two physical modes with non-trivial gamma phases, $
\gamma_l \ne 1.$   The version of  this rule derived
in~\cite{Kobayashi:2004ya}   uses instead the conjugacy classes of the lattice cosets,
$\L /\L _{\T ^g } $, associated to the fixed points $f ^l_{a _l} = {a _l\over 3}
e_1, \ [a _l=0,1,2] $ in sectors $g={2,4} $ and $f ^l_{a _l , b _l} = {a _l\over 2}
e_1 + {b _l\over 2} e_2 , \ [a _l=0,1; \ b _l=0,1]$ in sector $ g={3} $, 
in the  notational  conventions of Table~\ref{taba1}   and Eq.~(\ref{eqlatt3}).
In terms of the fixed point content of the physical states (with fixed
$\g _l$ eigenvalues),  the   orbifold space  group  constraints are  described
for the couplings  $  T_{2,4}^{p_1} T_0^{p_2}$  by the $Z_3 $ group selection rule, 
$ \sum _{l=1}^n a _l =0 \ \text{mod} \  3$, and for the   couplings  
$ T_{3} ^{p_1}T_0^{p_2}  $   by the $Z_2\times Z_2 $ group
selection rule, $ \sum _{l=1}^n a _l   =0 \ \text{mod} \  2 , \ \sum
_{l=1}^n b _l =0 \ \text{mod} \   2$. 
\end{itemize} 

Convincing arguments are given
in~\cite{Buchmuller:2006ik,Lebedev:2007hv,deGroot08} that Rule $I$ is
not a genuine selection rule, since it is found to be automatically
satisfied by physical states obeying the GSO orbifold projection, $P
(g, n_f, \g , \phi ) =1$, once the selection rules on $\T _I^{
g_l} , \ P ^I (l) , \ R^I(l) $ are imposed. In a consistent string
model complying with the orbifold GSO projection, and hence with the
world sheet modular group invariance, Rule $I$ is redundant.
However, the  redundancy  was proved with the specific
prescription for the orbifold projection quoted in Eq.(\ref{eqgso5}).
Since our prescription in~Eq.(\ref{eqgso3p}) is quite close but not
strictly identical to that used
in~\cite{Buchmuller:2006ik,Lebedev:2007hv}, there is no guarantee that
Rule $I$ is automatically implied by other rules in our case too. 
That  a selection rule  is superfluous in one  projection 
prescription and not in another   appears  odd at  first sight  but
cannot  be excluded. It  certainly is a logical possibility 
if several consistent string models   arise  from the same 
data set   of  shift vectors, $V,\ W_2,\ W_3 $.  To  settle this issue  on
practical grounds, we have applied our search
procedure for models $A1, \ B$ with the rule $\prod _{l=1}^n \g _l
=1$ included,  and found that this did forbid  
certain superpotential couplings allowed by the rules in
Eq.(\ref{eqsel}).   However, the excluded  couplings represent 
 a very tiny fraction  of the large  set of allowed couplings,  
which cause  insignificant changes on the effective couplings.
We have also  tested the  Rule II  above  and   found that this had a   
negligible incidence  on the allowed   couplings. 

%REFER TO TABLES!!!!!
%\ref{tabmoda1}  \ref{dirplatesa1} \ref{tabmdta2} \ref{tabmodeleb}
%\ref{dirplatb} \ref{taba1} \ref{taba2}

%%%%%%%%%%%%%%%%%%%FIN DE TEXTE%%%%%%%%%%%%%%%%%%%%

\end{document}